\documentclass[reprint, prx, superscriptaddress]{revtex4-2}

\usepackage{verbatim}
\usepackage{graphicx}
\usepackage{color}
\usepackage{amsmath}
\usepackage{amssymb}
\usepackage{braket}
\usepackage{enumerate}
\usepackage{multirow}
\usepackage{colortbl}
\definecolor{Gray}{gray}{0.3}
\usepackage{wrapfig}
\usepackage[latin1]{inputenc}

\DeclareMathAlphabet{\mathcal}{OMS}{cmsy}{m}{n}

\newcommand{\be}{\begin{equation}}
\newcommand{\ee}{\end{equation}}
\newcommand{\ti}[1]{\text{#1}}

\newcommand{\fig}[1]{Fig.~\ref{#1}}

\newcommand{\Fig}[1]{Figure~\ref{#1}}

\newcommand{\eq}[1]{Eq.~\eqref{#1}}
\newcommand{\eqs}[2]{Eqs.~\eqref{#1} and \eqref{#2}}
\newcommand{\Eq}[1]{Equation~\eqref{#1}}
\newcommand{\Eqs}[2]{Equations~\eqref{#1} and \eqref{#2}}
\newcommand{\stn}[1]{Sec.~\ref{#1}}

\newcommand{\bea}{\begin{eqnarray}}
\newcommand{\eea}{\end{eqnarray}}
 
\newcommand{\ba}{\begin{array}}
\newcommand{\ea}{\end{array}}

\newcommand{\bl}{\begin{flalign}}
\newcommand{\enl}{\end{flalign}}

\newcommand{\half}{\frac{1}{2}}

\newcommand{\tr}{\ti{Tr}}

\usepackage{makecell}

\begin{document}

\title{Analog Quantum Simulation of the Dynamics of Open Quantum Systems with Quantum Dots and Microelectronic Circuits}

\author{Chang Woo Kim}
\email{cwkim66@jnu.ac.kr}
\affiliation{
    Department of Chemistry, University of Rochester, Rochester, New York 14627, USA
    }
\affiliation{
    Department of Chemistry, Chonnam National University, Gwangju 61186, South Korea
    }
    
\author{John M. Nichol}
\affiliation{
    Department of Physics, University of Rochester, Rochester, New York 14627, USA
    }

\author{Andrew N. Jordan}
\affiliation{
    Department of Physics, University of Rochester, Rochester, New York 14627, USA
    }
\affiliation{
    Institute for Quantum Studies, Chapman University, Orange, California 92866, USA
    }
    
\author{Ignacio Franco}
\email{ignacio.franco@rochester.edu}
\affiliation{
    Department of Chemistry, University of Rochester, Rochester, New York 14627, USA
    }
\affiliation{
    Department of Physics, University of Rochester, Rochester, New York 14627, USA
    }

\date{\today}

\begin{abstract}
We introduce a general setup for the analog quantum simulation of the dynamics of open quantum systems based on semiconductor quantum dots electrically connected to a chain of quantum $RLC$ electronic circuits. The dots are chosen to be in the regime of spin-charge hybridization to enhance their sensitivity to the $RLC$ circuits while mitigating the detrimental effects of unwanted noise. In this context, we establish an experimentally realizable map between the hybrid system and a qubit coupled to thermal harmonic environments of arbitrary complexity  that enables the analog quantum simulation of open quantum systems. We assess the utility of the simulator by numerically exact emulations that indicate that the experimental setup can faithfully mimic the intended target even in the presence of its natural inherent noise. 
We further provide a detailed analysis of the physical requirements on the quantum dots and the $RLC$ circuits needed to experimentally realize this proposal that indicates that the simulator can be created with existing technology. The approach can exactly capture the effects of highly structured non-Markovian quantum environments typical of photosynthesis and chemical dynamics, and offer clear potential advantages over conventional and even quantum computation. The proposal opens a general path for effective quantum dynamics simulations based on semiconductor quantum dots.

\end{abstract}

\maketitle

\section{Introduction}

Open quantum systems refer to microscopic quantum coherent systems that are coupled to a quantum environment. Computing the dynamics of open quantum systems with high precision is a central challenge in chemistry, physics and quantum information science.\cite{Nielsen2011,Schlosshauer2007,Breuer2002}
This is because most quantum systems of physical interest are in interaction with an environment that modulates its physical properties and introduces decoherence and relaxation routes in the dynamics.\cite{Gu2018,Hu2018,Kim2021}

The challenge is particularly difficult when trying to capture the effects of environments of relevance in chemistry (such as solvent, vibrations, protein scaffolds, polymer matrices, and electromagnetic radiation) as needed, for example, to develop better organic solar cells\cite{DeSio2017,Nelson2018,Popp2019} or understand vital processes such as photosynthesis\cite{Kreisbeck2012,Blau2018,CardosoRamos2019} and vision.\cite{Hahn2000,Marsili2019,Chuang2021}
These environments can operate at disparate timescales with varying interaction strengths to the system (thus highly structured), remember the dynamical history of the system (thus non-Markovian), and lead to both energy relaxation, loss of quantum coherence and environment-mediated interactions (thus many-body).

Despite the success of a variety of approximate methods,\cite{Corrigall1971,Tanimura1990,Meyer1990,Sun1998,White2004,Tannor2007,Wang2015,Liu2018,Strathearn2018,Tamascelli2019,Tretiak2020} tackling this key problem exactly with conventional computers remains a formidable task  due to the exponential scaling of the computational cost with system size and bath complexity. Even for a quantum computer, currently available machines and algorithms have been shown to be able to capture the open quantum system dynamics of simple models,\cite{Wang2011,Dicandia2015,Hu2020,HeadMarsden2020} but are unable to deal with the structured environments encountered in chemistry. Accurately modeling this class of problems would require a large number ($\sim 10^5-10^7$) of entangled qubits,\cite{McArdle2020,Sawaya2020} which is a desirable but currently unreachable goal.


\begin{figure*}[htbp]
    \includegraphics[width=1.0\textwidth]{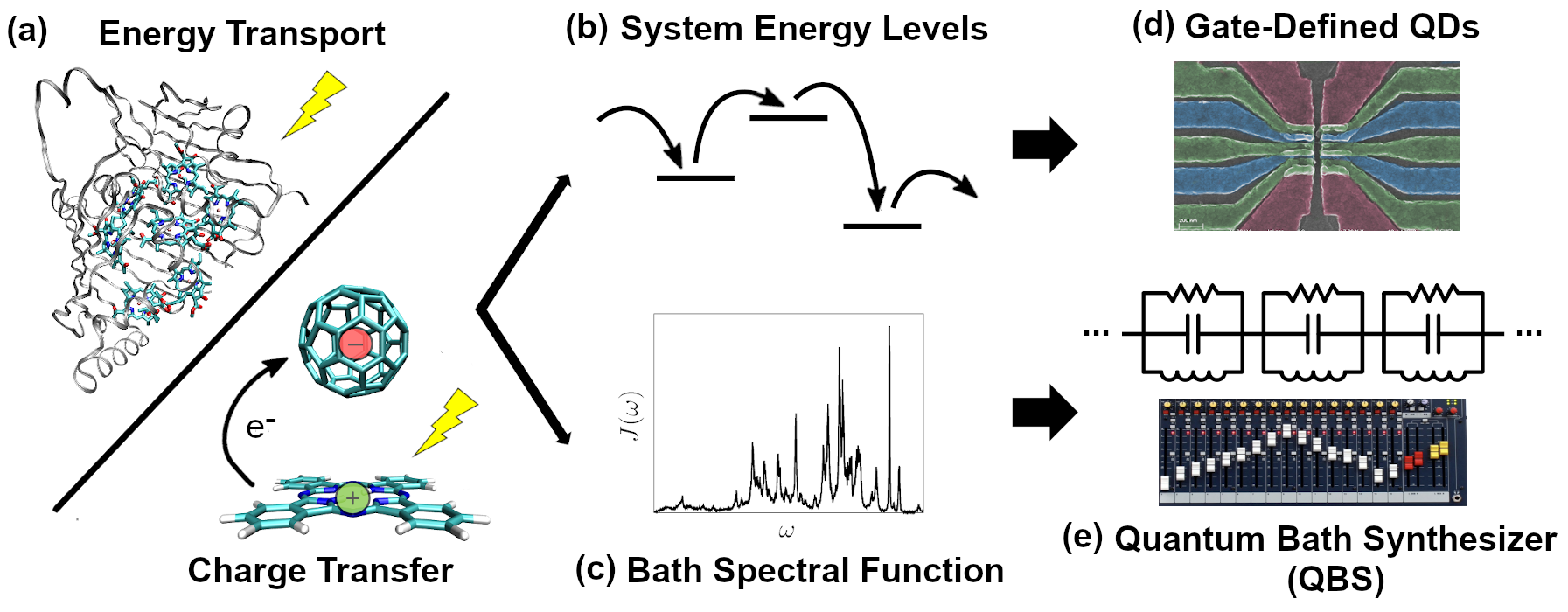}
    \caption{Schematic of the proposed quantum simulator of open quantum systems based on gate-defined QDs to model the system  and a Quantum Bath Synthesizer (QBS) composed of arrays of RLC circuits to model the bath. The proposed setup is versatile enough to capture environments of chemical complexity (a)-(b).}
    \label{fig:scheme}
\end{figure*}

An attractive alternative is analog quantum simulation,\cite{Cirac2012,Georgescu2014,Mostame2017,Kassal2021} which does not require a scalable quantum computer. 
In this approach, instead of trying to solve the Liouville-von Neumann equation using digital computation, the problem of physical interest is mapped onto a highly controllable experimental setup and nature is allowed to do the simulations. However, not all existing setups can be used for open quantum systems, partly because it is challenging to model the effect of the quantum environment on the system of interest. 
Current approaches in quantum simulation~\cite{Wang2018,Potocnik2018,Maier2019} mimic the environment by simply introducing classical noise. This approach can yield an apparent loss of coherence, but fails to capture energy relaxation through spontaneous emission and other unique quantum features of the environment and its dynamics.\cite{Aghtar2017,Gu2019} 
Interesting proposals to generate dissipation with classical noise\cite{Chenu2017} are only valid thus far for weak system-bath interactions.
Thus, there is a critical need to develop general methods to simulate the effects of complex quantum environments. 



In this paper, we develop the theory of a new simulator architecture for open quantum system dynamics harnessing semiconductor quantum dots and quantum electronic circuits, see \fig{fig:scheme}. We focus on environments that can be described by a collection of harmonic oscillators, such as photonic and phononic environments. This class of environments is widely applicable because any system-environment problem can be rigorously mapped onto a system linearly coupled to an oscillator environment, provided the interaction can be dealt with to second order in perturbation theory.\cite{Leggett1987}
Makri has further argued that this is a common situation in the thermodynamic limit when the system-environment coupling is diluted over a large number of degrees of freedom.\cite{Makri1999}

\begin{figure*}[htbp]
    \includegraphics[width=1.0\textwidth]{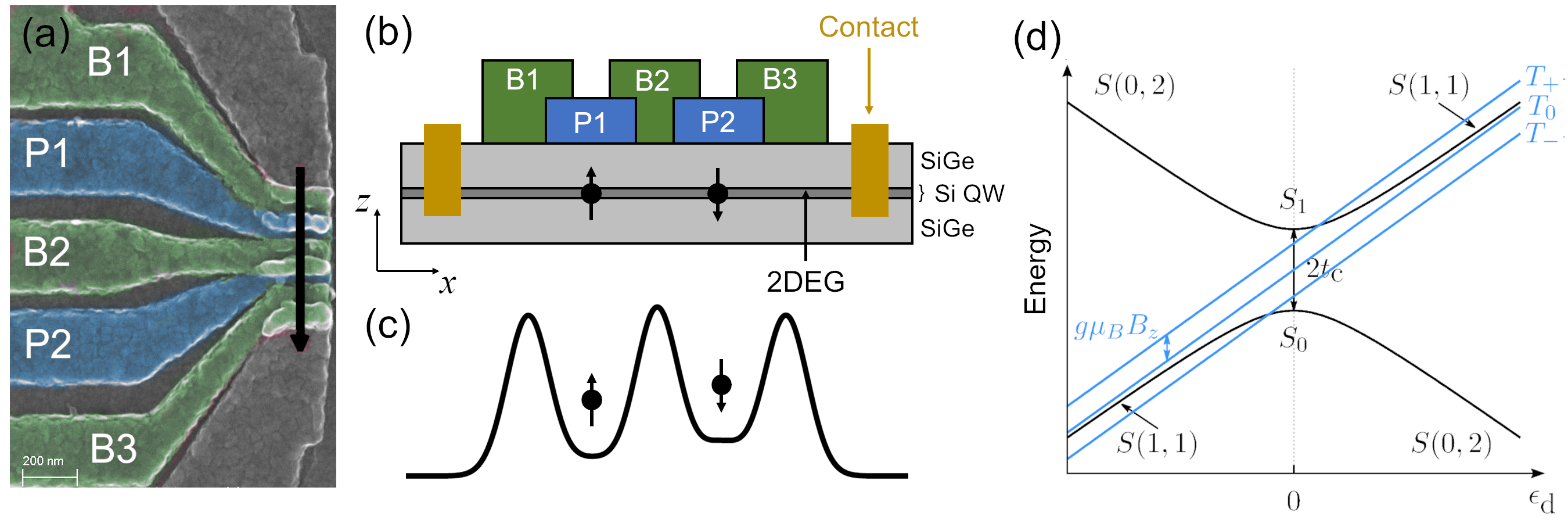}
    \caption{Semiconductor quantum dot setup. (a) False color scanning-electron micrograph of a double-quantum-dot (DQD) device formed by plunger (blue) and barrier (green) gates.
    (b) Scheme of the device and (c) electrostatic potential energy surface along the current path [the black arrow in (a)]. (d) Eigenenergies of the DQD with 2 electrons for a given detuning $\epsilon_\ti{d}$ and magnetic field $B_z$.}
    \label{fig:QD}
\end{figure*}

To model the system we propose the use of gate-defined semiconductor quantum dots (QDs)~\cite{Reimann2002,Hanson2007} as they enable the design of highly configurable and coherent quantum systems. These QDs are created by
lithographically fabricating nanoscale electrodes on the surface of semiconductor interfaces (e.g. GaAs/AlGaAs or Si/SiGe) hosting a two-dimensional electron gas at cryogenic temperatures. Through applied voltages on the electrodes, one can create electrostatic potential wells that act as QDs for electrons (\fig{fig:QD}) with a well-defined spectrum of discrete electronic energy levels. Through gate voltages, the shape and depth of the QDs, the tunnel coupling
between consecutive QDs, and the number of confined electrons can be precisely manipulated (\fig{fig:QD}b-c). State-of-the-art experiments using QDs involve the manipulation of up to 9 QDs in a linear geometry.\cite{Mills2019, Philips2022}
Recent experimental achievements include quantum teleportation between distant electron spins,\cite{Qiao2020} and rapid shuttling of single electron across a large QD array,\cite{Mills2019} all demonstrating the high controllability of the QD platform. These technical advances have positioned the QDs as an excellent candidate for analog quantum simulation. In fact, QDs have been proposed as simulators of simple chemical reactions~\cite{Smirnov2007} and used to simulate strongly correlated electron systems.\cite{Barthelemy2013, Hensgens2017}
Our proposed QD-based scheme to simulate open quantum systems complement these efforts and has unique advantages over related proposals based on superconducting qubits\cite{Mostame2017,Leppaekangas2018} and ion traps~\cite{Kassal2021}.

To model the environment we introduce the concept of a \emph{quantum bath synthesizer} (QBS) that is composed of arrays of quantum electronic circuits (\fig{fig:scheme}e). The approach is based on cooling $RLC$ circuits until they behave like dissipative quantum mechanical oscillators. By judiciously controlling the resistance ($R$), inductance ($L$), and capacitance ($C$) of the circuits, the frequency, quantum fluctuations and relaxation of each oscillator can be tailored. By considering an array of them with different frequencies, the physical system will act as a quantum bath that can be tuned to mimic the dynamics and response of thermal environments even of chemical complexity. Using the $RLC$ degrees of freedom gives an analog quantum simulator an advantage over digital approaches which require many qubits to accurately represent a single harmonic oscillator. 

The strategy to use quantum circuits to introduce dissipation in analog simulation was first suggested by Mostame \emph{et al.}\cite{Mostame2012,Mostame2017} for flux qubits inductively coupled to $RLC$ circuits. In this proposal, however, the physical proximity required for effective inductive coupling limits the number of distinct circuit elements that can be  coupled to the qubit. Lepp\"akangas \emph{et al.}\cite{Leppaekangas2018} adapted the idea for transmons electrically coupled to microwave resonators and proposed a scheme based on two-color driving to reproduce Ohmic environments. However, the physical requirements needed to capture environments of chemical complexity remain unclear. 
Our proposal can be used to model complex environments and is based on purely electrical interactions between the QDs and the $RLC$ circuits. These electrical connections offer fast response times and the ability to spatially separate the QDs from the QBS offering additional flexibility in designing the simulator and controlling its parameters.


The proposed approach constitutes a general modeling strategy for open quantum systems based on QDs that can be used to understand the operation of realistic quantum devices, to engineer quantum environments that enhance system function, to isolate qubits with enhanced coherence properties, to understand elementary steps in photosynthesis, and to test quantum control strategies in the presence of quantum environments. As in conventional simulation, the setup enables continuous tuning of the Hamiltonian through external controls opening essential routes to understand properties of matter that are not accessible to direct experimentation. In addition it has the advantage that the computational cost does not increase with the complexity of the spectral density or in regimes in which the quantum features of the environment become increasingly important. The proposed strategy captures dissipative effects to all orders in the system-bath coupling and includes both Markovian and non-Markovian effects.

Specifically, in this paper we provide a strategy to build an analog quantum simulator based on QDs for a qubit coupled to a thermal environment of arbitrary complexity. In \stn{stn:mapping0} we develop a rigorous mapping between the target system and the quantum simulator. In \stn{stn:emulations} we use numerically exact emulations to examine simulation accuracy even in the presence of experimental noise. These emulations reveal clear operation conditions where the simulator can accurately mimic the target dynamics. In \stn{stn:experimentalneeds} we analyze the experimental requirements needed for its realization and find that building such a simulator is accessible to present-day technology. Last, in \stn{stn:extensions}, we discuss how the proposed simulator can be used as a building block for the general simulation of open quantum systems. We summarize our observations in \stn{stn:conclusions}.


\section{Mapping open quantum systems to hybrid quantum dots-QBS systems}\label{stn:mapping0}

In developing a useful analog simulator, the key is to isolate degrees of freedom of the QDs that offer customizable energy level structure and interact strongly with the QBS but weakly with the natural environment (i.e. the charge fluctuations and nuclear spins in the semiconductor) such that the simulator informs about the system of interest. At the bath level, the key is to guarantee that the QBS is customizable such that it can be used to model a wide range of environments and that the interactions between the QDs and the QBS  accurately map into the physical interactions of interest.
The QDs offer charge, spin and hybrid spin-charge (singlet/triplet configurations of electron pairs) controllable degrees of freedom. Here we hypothesize that hybrid spin-charge approaches are likely the most suitable, as they avoid the short coherence time of charge, and the weak interactions of spin. 

\subsection{Target Hamiltonian}\label{subsection:H_orig}
We consider a general two-level quantum system that consists of a ground state $\ket{g}$ and an excited state $\ket{e}$ in interaction with a thermal bath. The target Hamiltonian that we want to encode in the quantum simulator is of the form
\begin{equation}\label{eq:H_full}
    \hat{H} = \hat{H}_\ti{s} \otimes \hat{\mathbf{1}}_\ti{b} + \hat{\mathbf{1}}_\ti{s} \otimes \hat{H}_\ti{b} + \hat{H}_\ti{int},
\end{equation}
where $\hat{H}_\ti{s}$ describes the system part, $\hat{H}_\ti{b}$ the bath and $\hat{H}_\ti{int}$ their interaction, and $\hat{\mathbf{1}}_\ti{s}$ ($\hat{\mathbf{1}}_\ti{b}$) is the identity operator in the system (bath) subspace.
 The general Hamiltonian for the qubit is 
\begin{equation}\label{eq:H_subsys}
    \hat{H}_\ti{s} = \frac{\Delta}{2} \hat{\sigma}_z + \eta \hat{\sigma}_x,
\end{equation}
where $\Delta$ is the energy difference between the two states, $\eta$ their coupling, and $\hat{\sigma}_x = \ket{e}\bra{g} + \ket{g}\bra{e}$ and $\hat{\sigma}_z = \ket{e}\bra{e} - \ket{g}\bra{g}$ the Pauli spin operators. In turn, the bath consists of a collection of harmonic oscillators,
\begin{equation}\label{eq:H_bath}
    \hat{H}_\ti{b} = \sum_j \hbar \omega_j \bigg( \hat{a}_j^\dagger \hat{a}_j + \frac{1}{2} \bigg),
\end{equation}
where $\hat{a}_j^\dagger$ and $\hat{a}_j$ are the raising and lowering operators of the $j$-th mode with characteristic frequency $\omega_j$. 
For the system-bath interaction,  we consider the archetypical spin-boson and displaced harmonic oscillator (or Frenkel-Holstein) models as they have been found to be useful to describe a variety of problems. This includes
quantum impurity problems,\cite{Guinea1985} quantum thermodynamics\cite{Newman2017}, artificial light-matter coupling\cite{FriskKockum2019} for the spin-boson coupling, and molecular photophysics such as the dynamics of natural\cite{Jang2018} and artificial chromophore aggregates,\cite{Hestand2018} photo-voltaic materials,\cite{Janke2020} and electro-luminescent materials\cite{Berkelbach2013} for the displaced harmonic oscillator. 

In both cases, the  displacement of the collective bath coordinate 
\be\label{eq:op_B}
\hat{B}= \sum_j \hbar g_j (\hat{a}_j^\dagger + \hat{a}_j),
\ee
leads to fluctuations in the system energy. That is, $\hat{H}_\ti{int}= \hat{S} \otimes \hat{B}$ where $g_j$ is the coupling strength of the $j$-th mode to the system. In the spin-boson model $\hat{S}=\hat{\sigma}_z$ and the ground and excited states are symmetrically affected. In turn, in the displaced oscillator model  only the excited state is affected and $\hat{S}= \ket{e}\bra{e}$. The influence of the harmonic bath on the system is completely characterized by its spectral density,\cite{Leggett1987}
\begin{equation}\label{eq:BSD}
    J(\omega) = \sum_j \hbar g_{j}^2\delta(\omega - \omega_j),
\end{equation}
a quantity that summarizes the frequencies of the environment and the system-bath coupling strengths. 

We suppose that at $t = 0$, the density matrix of the composite system is factorizable $\hat{\rho}(0) = \hat{\rho}_\ti{s}(0) \otimes \hat{\rho}_\ti{b}(0)$, where $\hat{\rho}_\ti{s}(t)$ is the reduced density matrix of the system and $\hat{\rho}_\ti{b}(t)$ that of the bath. At initial time, the bath is taken to be at thermal equilibrium at temperature $T$, i.e.
\be \label{eq:Bath_DM}
    \hat{\rho}_\ti{b}(0) = \frac{ \exp(-\hat{H}_\ti{b}/k_\ti{B} T) }{ \tr[ \exp(-\hat{H}_\ti{b}/k_\ti{B} T) ]}.
\ee

The feasibility of the analog quantum simulation relies on the ability to experimentally construct artificial systems that faithfully mimic $\hat{H}_\ti{s}$, $J(\omega)$ and the initial conditions. While the dynamics of interest often operate around room temperature $T \sim 300$ K, the QDs operate at cryogenic temperatures $T_\ti{qs} < 1$ K.
To establish a map, energies and timescales need to be scaled according to the temperature ratio 
\be
\label{eq:gamma}
\gamma= \frac{T}{T_\ti{qs}}.
\ee 
The dynamics of the simulator will be $\gamma$ times slower, and the characteristic energies $1/\gamma$ smaller than in the physical system. Given a useful mapping, this guarantees that the effective dynamics and thermal density matrices of the simulator and its target coincide. This temperature scaling factor can be used to map a wide range of target Hamiltonians to the setup and offers the additional advantage of slowing down and thus increasing the time resolution of the simulation with respect to direct experimentation in the target system.

\subsection{Mapping thermal environments to arrays of \emph{RLC} circuits}
\label{stn:pre_qbsdesign}

To emulate thermal environments we consider arrays of $RLC$ circuits. The elementary unit is a parallel $LC$ circuit containing a coil with inductance $L$ (the ability to oppose a change in the electric current arising from electromagnetic induction) and a capacitor with capacitance $C$ (the ability to store charge at a given voltage). The total classical Hamiltonian for the $LC$ circuit is
\begin{equation} \label{eq:H_LC}
    H_{LC} = \frac{Q^2}{2C} + \frac{\Phi^2}{2L},
\end{equation}
where $Q$ is the charge stored in the capacitor and $\Phi$ is the magnetic flux passing through the inductor. \Eq{eq:H_LC} is identical to the Hamiltonian of a harmonic oscillator of frequency $\Omega_0 = 1/\sqrt{LC}$ and mass $m = C$ if we simply identify the momentum $p \rightarrow Q$ and position $q \rightarrow \Phi$. In fact, just as with $p$ and $q$, $Q$ and $\Phi$ are canonically conjugated variables \cite{Devoret1997}. That is, the corresponding quantum mechanical operators $\hat{Q}$ and $\hat{\Phi}$ satisfy $[\hat{\Phi}, \hat{Q}] = i\hbar$. This allows us to re-write \eq{eq:H_LC} in quantized form as
\begin{equation}\label{eq:H_LC_oper}
    \hat{H}_{LC} = \hbar \Omega_0 \bigg( \hat{b}^\dagger \hat{b} + \frac{1}{2} \bigg),
\end{equation}
where $\hat{b}^\dagger$ and $\hat{b}$ are the raising and lowering operators defined as
\begin{equation}\label{eq:LC_ladder_oper}
    \begin{gathered}
    \hat{b}^\dagger = \sqrt{\frac{1}{2\hbar Z_0}}(Z_0 \hat{Q} + i \hat{\Phi}), \\
    \hat{b} = \sqrt{\frac{1}{2\hbar Z_0}}(Z_0 \hat{Q} - i \hat{\Phi}),
    \end{gathered}
\end{equation}
with $Z_0 = \sqrt{L/C}$ the characteristic impedance. For low temperatures satisfying $k_\ti{B}T_\ti{qs} \ll \hbar\Omega_0$, the $LC$ circuit behaves exactly like a quantum harmonic oscillator and the electromagnetic observables $\hat{Q}$ and $\hat{\Phi}$  show zero-point quantum fluctuations.  Fluctuations in the charge of the $LC$ circuit introduce fluctuations in the circuit voltage $\hat{V}_{LC}= \hat{Q}/C $ that influence the potential wells of the QDs. These zero-point fluctuations of circuit voltage form the physical basis of the system-bath coupling in the simulator.

A rigorous analogy between thermal environments and an array of quantum $RLC$ circuits is established as follows. The QBS needs to reproduce spectral densities encountered in thermal environments. To understand the mapping between these two, it is useful to focus on the two-time bath correlation function~\cite{Schlosshauer2007}
\begin{multline}\label{eq:TCF_0}
    \langle \hat{B}(t)\hat{B}(0) \rangle_{T} =\\
    \hbar \int_0^\infty J(\omega) \bigg[ \coth \bigg( \frac{\hbar\omega}{2k_\ti{B}T} \bigg)\cos\omega t - i\sin \omega t \bigg] d\omega,
\end{multline}
where the angular bracket $\langle \cdots \rangle_T$ denotes a thermal average at temperature $T$, and $\hat{B}(t)$ are the bath operators in $\hat{H}_\ti{int}$ [\eq{eq:op_B}] in the interaction picture of $\hat{H} - \hat{H}_\ti{int}$.
\Eq{eq:TCF_0} is a useful quantity for describing harmonic environments because 
all higher-order time correlations can be written in terms of \eq{eq:TCF_0} in this case.\cite{Weiss1999,Chen2017} We now use it to identify the proper mapping.

Consider a QBS made of a single $LC$ circuit  coupled to the QDs through voltages $\hat{V}_{LC}= \hat{Q}/C $ that influence the potential wells, i.e. $\hat{H}_\ti{int}^\ti{qs} = \hat{S}_\ti{qs} \otimes \hat{B}_\ti{qs}$ with $\hat{B}_\ti{qs} = \kappa \hat{V}_{LC}$, where $\kappa$ is a response factor that quantifies how the energy of a QD changes with applied $LC$ voltage. 
Since the $LC$ circuit behaves like a harmonic oscillator, the time-correlation function for $\hat{B}_\ti{qs}$ is~\cite{Devoret1997}
\begin{multline}\label{eq:TCF_LC}
    \langle \hat{B}_\ti{qs}(t) \hat{B}_\ti{qs}(0) \rangle_{T_\ti{qs}} =  \\ \frac{\hbar \Omega_0^2 Z_0}{2} \kappa^2  \bigg[\coth \bigg( \frac{\hbar \Omega_0}{2k_\ti{B} T_\ti{qs}} \bigg) \cos\Omega_0 t - i\sin \Omega_0 t \bigg].
\end{multline}
We now introduce the impedance $Z(\Omega)$, a complex-valued quantity which describes the response of a circuit under a sinusoidal voltage  of frequency $\Omega$. The magnitude $|Z(\Omega)|$ is a resistance while the argument $\arg[Z(\Omega)]$ describes the phase difference between the voltage and current. The impedance of an $LC$ circuit is~\cite{Devoret1997}
\begin{multline}\label{eq:LC_impedance}
    Z(\Omega) = \frac{\pi \Omega_0 Z_0}{2}[\delta(\Omega-\Omega_0) + \delta(\Omega+\Omega_0)] \\
    + \frac{iZ_0}{2}\bigg[\mathcal{P}\bigg(\frac{\Omega_0}{\Omega-\Omega_0}\bigg) + \mathcal{P}\bigg(\frac{\Omega_0}{\Omega+\Omega_0}\bigg)\bigg],
\end{multline}
where $\mathcal{P}$ denotes the principal value. From \eqs{eq:TCF_LC}{eq:LC_impedance}, it follows that 
\begin{multline}\label{eq:TCF_qs}
    \langle \hat{B}_\ti{qs}(t) \hat{B}_\ti{qs}(0) \rangle_{T_\ti{qs}} = \frac{\hbar \kappa^2}{\pi} \int_0^\infty \Omega \: \ti{Re} [Z(\Omega)] \\
    \times \bigg[ \coth \bigg( \frac{\hbar \Omega}{2k_\ti{B} T_\ti{qs}} \bigg) \cos \Omega t - i\sin \Omega t \bigg] d\Omega.
\end{multline}
To mimic \eq{eq:TCF_0}, one needs to consider a collection of serially connected $LC$ circuits with different values of $L$ and $C$, with Hamiltonian
\begin{equation}\label{eq:H_QBS}
    \hat{H}_\ti{QBS} = \sum_j \hbar \Omega_j \bigg( \hat{b}_j^\dagger \hat{b}_j + \frac{1}{2} \bigg).
\end{equation}
where the bosonic ladder operators $\hat{b}_j^\dagger$ and $\hat{b}_j$ are defined analogously to \eq{eq:LC_ladder_oper},
\begin{equation}\label{eq:LC_ladder_j}
    \begin{gathered}
    \hat{b}_j^\dagger = \sqrt{\frac{1}{2\hbar Z_{0j}}}(Z_{0j} \hat{Q}_j + i \hat{\Phi}_j), \\
    \hat{b}_j = \sqrt{\frac{1}{2\hbar Z_{0j}}}(Z_{0j} \hat{Q}_j - i \hat{\Phi}_j).
    \end{gathered}
\end{equation}
Here, $Z_{0j}$ is $Z_0$ for the $j$-th oscillator. The continuous limit can be achieved with a finite number of serially connected parallel $RLC$ circuits where the resistance broadens their spectral features and introduces dissipation (see \fig{fig:scheme}e). The validity of \eq{eq:TCF_qs} remains unaffected at this limit, but the impedance of the serial circuit is now\cite{Devoret1997}
\begin{equation}\label{eq:RLC_SPD}
    \begin{gathered}
    \text{Re}[Z(\Omega)] = \sum_{j=1} \text{Re}[Z_j(\Omega)], \\
    \text{Re}[Z_j(\Omega)] = 2\Lambda_j \frac{2\Gamma_j\Omega^2}{(\Omega_j^2 - \Omega^2)^2 + 4\Gamma_j^2 \Omega^2},
    \end{gathered}
\end{equation}
where $Z_j(\Omega)$ is the contribution of a single $RLC$ circuit. The reorganization energy $\Lambda_j = Z_{0j} \Omega_j / 2$ and damping constant $\Gamma_j = Z_{0j} \Omega_j / 2R_j$ can be controlled by varying the circuit resistance $R_j$, frequency  $\Omega_j$ and $Z_{0j} = \sqrt{L_j / C_j}$. For $R_j \to 0$, the contribution of the circuit to the spectral density vanishes as $\ti{Re}[Z_j(\Omega)] \to 0$. For finite $R_j$, the resistance introduces broadening of the resonance of the $LC$ circuit. As $R_j\to \infty$, the $RLC$ circuit reduces to an $LC$ circuit and $\Gamma_j \to 0$.

Comparing \eqs{eq:TCF_0}{eq:TCF_qs}, we see that the $RLC$ circuits behave identically to the target environment provided one tailors $Z(\Omega)$ such that
\begin{equation}\label{eq:spd_vs_RLC}
    J(\omega) = \frac{\kappa^2\omega}{\pi}\:\ti{Re}[Z(\omega/\gamma)],
\end{equation}
where $\gamma$ is the temperature ratio in \eq{eq:gamma}. This equation is central to establish an exact map between thermal environments and $RLC$ circuits.



\subsection{Mapping system levels to QD states}
\label{stn:mapping}

We now address the key issue of how to map the qubit Hamiltonian [\eq{eq:H_subsys}] into the QD states. For definitiveness, we focus on the double quantum dot (DQD) with 2 electrons schematically shown in \fig{fig:QD}.  To proceed it is important to understand the energy level structure offered by the DQD,\cite{Petta2005,Foletti2009,Wong2015}  schematically shown in \fig{fig:QD}d. Here, $(1,1)$ describes the state in which one electron occupies both the left and right dots and (0,2) in which both electrons occupy the right dot. The energy of these states are $\epsilon_{(1,1)} = \epsilon_\ti{L}+\epsilon_\ti{R}$ and $\epsilon_{(0,2)} = 2\epsilon_\ti{R} + \epsilon_\ti{pair}$,
where $\epsilon_{\ti{L}/\ti{R}}$ denotes the ground state energy of the left/right QD and $\epsilon_\ti{pair}$ is the electron pairing energy arising from Coulombic interactions. The energetic detuning $\epsilon_\ti{d}$ ($x$-axis in the \fig{fig:QD}d) is defined as the energy difference between the two charge configurations,
\begin{equation}\label{eq:detuning}
    \epsilon_\ti{d} \equiv \epsilon_{(1,1)} - \epsilon_{(0,2)} = \epsilon_\ti{L} - \epsilon_\ti{R} - \epsilon_\ti{pair}.
\end{equation}
In the absence of a magnetic field, the (1,1)  configuration is four-fold degenerate with one singlet and three triplet spin states. By contrast, the (0,2) configuration favors the singlet state due to the Pauli exclusion principle.

The singlet state of (1,1) [$\ket{S(1,1)}$] and (0,2) [$\ket{S(0,2)}$] configurations hybridize near the degeneracy point $\epsilon_\ti{d} \sim 0$ due to the tunnel coupling $t_\ti{c}$ between the QDs, yielding two adiabatic singlet states [$\ket{S_0}$, $\ket{S_1}$]. By contrast, the (1,1) triplet states [$\ket{T_{0, \pm}(1,1)}$ or $\ket{T_{0,\pm}}$ for short] are not affected by the hybridization. 

Applying an external magnetic field along the $z$-axis lifts the energy degeneracy of triplet states. The energy diagram in \fig{fig:QD}d results from these observations, where the zero-energy baseline is chosen as $(\epsilon_\ti{L} + 3\epsilon_\ti{R}+ \epsilon_\ti{pair})/2$ to make the diagram symmetric. In the idealized picture, the $\epsilon_{\ti{L}/\ti{R}}$ are manipulated by changing the gate potential of the left/right dot (P1 and P2 in~\fig{fig:QD}b) while the tunnel coupling $t_\ti{c}$ is controlled through barrier-gate electrodes between the QDs (B2 in~\fig{fig:QD}b). The resulting level mixing is referred to as spin-charge hybridization. 

In mapping the open quantum system to this hybrid structure there is freedom on which QD states to use. However, the choice can dramatically impact the experimental requirements and utility of the simulator. We now introduce a useful mapping for two-state problems based on the $S_0/T_-$ states. 
These states are chosen because their relative energy is controllable by varying $\epsilon_\ti{d}$ (\fig{fig:QD}c), their coupling can be manipulated through magnetic fields, they are not highly excited states that suffer from additional spontaneous relaxation, and because they can be used to form useful qubits.~\cite{Nichol2021} In this map, the quantum noise is incorporated by connecting the QBS to the left QD, leading to  quantum fluctuations in the $\epsilon_{\ti{L}}$ and thus $\epsilon_\ti{d}$.

Other alternative mappings suffer from significant drawbacks.  For instance, choosing states that are distinguished solely by charge configurations such as $S(1,1)$ and $S(0,2)$ suffer from the fast unwanted decoherence due to coupling to charge fluctuations in the natural environments of the QDs, making them of limited utility for simulation. In turn, states that differ in spin but not charge, such as $S_0$ and $T_0$ for negative detunings,  offer long coherence times as they are mostly uncoupled to charge fluctuations in the natural QD environment but have the disadvantage of also being insensitive to the desired charge fluctuations introduced by the QBS, making the experimental requirements on the QBS unfeasible.

The Hamiltonian for a DQD in the presence of an external magnetic field is $\hat{H}_\ti{QD} = \hat{H}_\ti{QD}^0 + g\mu_B(\mathbf{B}_\ti{L} \cdot \mathbf{S}_\ti{L} + \mathbf{B}_\ti{R} \cdot \mathbf{S}_\ti{R})$,
where $\hat{H}_\ti{QD}^0$ is the pristine quantum dot Hamiltonian, $g$ the Land\'e $g$-factor, $\mu_B$ the Bohr magneton,  $\mathbf{B}_{\ti{L}/\ti{R}}$ the magnetic field, and $\mathbf{S}_{\ti{L}/\ti{R}}$ the total spin operator of the electron(s) at the left/right QD.
We take the magnetic field to have the form of $\mathbf{B}_\ti{L/R} = \pm \frac{\Delta B}{2}\hat{x} - B_\ti{avg}\:\hat{z}$, such that it changes between the dots (see \fig{fig:QD}b). In the \{$\ket{T_+}$, $\ket{T_0}$, $\ket{T_-}$, $\ket{S(1,1)}$, $\ket{S(0,2)}$\} basis, 
\begin{widetext}
\begin{equation}\label{eq:Hqd_unhyb}
    \hat{H}_\ti{QD} = \frac{\epsilon_\ti{L} + 3\epsilon_\ti{R}+ \epsilon_\ti{pair}}{2} \hat{\mathbf{1}}_\ti{QD} 
    + \begin{pmatrix}
    \frac{\epsilon_\ti{d}}{2} & 0 & 0 & 0 & 0\\
    0 & \frac{\epsilon_\ti{d}}{2} & 0 & 0 & 0\\
    0 & 0 & \frac{\epsilon_\ti{d}}{2} & 0 & 0\\
    0 & 0 & 0 & \frac{\epsilon_\ti{d}}{2} & t_\ti{c}\\
    0 & 0 & 0 & t_\ti{c} & -\frac{\epsilon_\ti{d}}{2}
    \end{pmatrix} + g\mu_B \begin{pmatrix}
    B_\ti{avg} & 0 & 0 & -\frac{\Delta B}{2\sqrt{2}} & 0 \\
    0 & 0 & 0 & 0 & 0\\
    0 & 0 & -B_\ti{avg} & \frac{\Delta B}{2\sqrt{2}} & 0\\
    -\frac{\Delta B}{2\sqrt{2}} & 0 & \frac{\Delta B}{2\sqrt{2}} & 0 & 0 \\
    0 & 0 & 0 & 0 & 0
    \end{pmatrix},
\end{equation}
\end{widetext}
where $\hat{\mathbf{1}}_\ti{QD}$ is the identity operator in the QD Hilbert space.
From this point on we disregard the zero base line of energy as it does not change the dynamics. 

The desired mapping is obtained by expressing the Hamiltonian in the eigenbasis of  $\hat{H}_\ti{QD}^0$, \{$\ket{T_+}$, $\ket{T_0}$, $\ket{T_-}$, $\ket{S_0}$, $\ket{S_1}$\}, and focusing on the $\ket{T_-}$, $\ket{S_0}$ subspace. In this basis,
\begin{widetext}
\begin{equation}\label{eq:simulator_bare}
    \hat{H}_\ti{QD} = \begin{pmatrix}
    \frac{\epsilon_\ti{d}}{2} & 0 & 0 & 0 & 0\\
    0 & \frac{\epsilon_\ti{d}}{2} & 0 & 0 & 0\\
    0 & 0 & \frac{\epsilon_\ti{d}}{2} & 0 & 0\\
    0 & 0 & 0 & \epsilon_{S_0} & 0 \\
    0 & 0 & 0 & 0 & \epsilon_{S_1}
    \end{pmatrix} +g\mu_B \begin{pmatrix}
    B_\ti{avg} & 0 & 0 & -\frac{\Delta B}{2\sqrt{2}}\sin\theta & -\frac{\Delta B}{2\sqrt{2}}\cos\theta \\
    0 & 0 & 0 & 0 & 0 \\
    0 & 0 & -B_\ti{avg} & \frac{\Delta B}{2\sqrt{2}}\sin\theta & \frac{\Delta B}{2\sqrt{2}}\cos\theta \\
    -\frac{\Delta B}{2\sqrt{2}}\sin\theta & 0 & \frac{\Delta B}{2\sqrt{2}}\sin\theta & 0 & 0 \\
    -\frac{\Delta B}{2\sqrt{2}}\cos\theta & 0 & \frac{\Delta B}{2\sqrt{2}}\cos\theta & 0 & 0
    \end{pmatrix}.
\end{equation}
\end{widetext}
The $\ket{S_0}$ and $\ket{S_1}$ states are defined by  
\be\label{eq:sc_hyb}
\begin{gathered}
    \ket{S_0} \equiv \sin\theta \ket{S(1,1)} - \cos\theta \ket{S(0,2)},\\
    \ket{S_1} \equiv \cos\theta \ket{S(1,1)} + \sin\theta \ket{S(0,2)},
\end{gathered}
\ee
with the mixing angle
\be\label{eq:mixingangle}
    \theta = \left\{
        \begin{array}{cc}
            \displaystyle \frac{1}{2} \tan^{-1} \bigg( \frac{2t_\ti{c}}{\epsilon_\ti{d}} \bigg), & \epsilon_\ti{d} \ge 0, \\
            \displaystyle \frac{\pi}{2} + \frac{1}{2} \tan^{-1} \bigg( \frac{2t_\ti{c}}{\epsilon_\ti{d}} \bigg), & \epsilon_\ti{d} < 0,
        \end{array}
    \right.
\ee
while their eigenenergies are $\epsilon_{S_{0/1}} = \mp \sqrt{\frac{\epsilon_\ti{d}^2}{4} + t_\ti{c}^2}$. In the  $\ket{T_-}$, $\ket{S_0}$  subspace this yields
\begin{multline}\label{eq:2state_qs}
  \hat{H}_\ti{QD}^\ti{qs} = E_0 (\ket{T_-}\bra{T_-} + \ket{S_0}\bra{S_0}) +\\ \Delta_\ti{qs}(\ket{S_0}\bra{S_0} - \ket{T_-}\bra{T_-}) + \eta_{\ti{qs}}(\ket{T_-}\bra{S_0} + \ket{S_0}\bra{T_-})
\end{multline}
where $E_0 = \frac{\epsilon_\ti{d}}{4} + \frac{\epsilon_{S_0} - g \mu_B B_\ti{avg}}{2}$ is the zero of energy, and the Hamiltonian parameters $\Delta_\ti{qs} = \epsilon_{S_0} - \frac{\epsilon_\ti{d}}{2} - g \mu_B B_\ti{avg}$ and $\eta_\ti{qs} = \frac{g \mu_B \Delta B}{2\sqrt{2}} \sin \theta$ are experimentally controllable.  
The ``qs'' superscript highlights that this is the subspace where the quantum simulation takes place. This desired subspace can be energetically isolated from the two remaining triplet states by increasing $B_\ti{avg}$, and from the $\ket{S_1}$ state by increasing the magnitude of the detuning. By simply changing the zero of energy, this yields the Hamiltonian in \eq{eq:H_subsys} by associating $\ket{T_-}$ with $\ket{g}$ and $\ket{S_0}$ with $\ket{e}$.

In the absence of the QBS,  $\epsilon_\ti{L/R}$ is purely determined by gate voltages, i.e. $\epsilon_\ti{L/R} =\epsilon_\ti{L/R}^g$ where $\epsilon_\ti{L/R}^g$ are energetic changes due to applied voltages on the lithographic electrodes. As discussed in \stn{stn:mapping2}, connecting the QDs to the QBS leads to additional static energy contributions to $\epsilon_\ti{L/R}$.

\subsection{Mapping the system-bath coupling}\label{stn:mapping2}

The total Hamiltonian of the QD-QBS hybrid system is of the form
\begin{equation}\label{eq:H_QD_QBS_hybrid}
    \hat{H}_\ti{exp} = \hat{H}_\ti{QD} \otimes \hat{\mathbf{1}}_\ti{QBS} + \hat{\mathbf{1}}_\ti{QD} \otimes \hat{H}_\ti{QBS} + \hat{H}_\ti{QD-QBS},
\end{equation}
where $\hat{H}_\ti{QD-QBS}$ describes the interactions between the QDs and the QBS. 
We now consider how to map the physical interaction between system and environment into that of the QDs with the QBS. 
Since the energetics of the QDs depends on $\epsilon_\ti{L} - \epsilon_\ti{R}$ [see \eq{eq:detuning}], for the QBS to exert quantum noise, the two dots must couple differently to the QBS. To be concrete, we choose to couple the QBS to the left QD only 
\begin{equation}\label{eq:qb_qds}
\vspace{-0.15cm}
    \hat{H}_\ti{QD-QBS} =  (\hat{\mathbf{1}}_\ti{QD} - \ket{S(0,2)} \bra{S(0,2)})\otimes \alpha \sum_{j} \hat{V}_j,
\vspace{-0.15cm}
\end{equation}
where $\hat{V}_j = \hat{Q}_j / C_j$ is the voltage operator for the $j$-th $LC$ circuit of the QBS and $\alpha$ is the ``lever arm'' or proportionality factor which converts changes in gate voltage to changes in the QD energy. In the \{$\ket{T_+}$, $\ket{T_0}$, $\ket{T_-}$, $\ket{S_0}$, $\ket{S_1}$\} basis:
\begin{widetext}
\begin{equation}\label{eq:simulator_hybrid}
     \hat{H}_\ti{QD-QBS} = \begin{pmatrix}
     1 & 0 & 0 & 0 & 0\\
     0 & 1 & 0 & 0 & 0\\
     0 & 0 & 1 & 0 & 0\\
     0 & 0 & 0 & \sin^2 \theta & \sin\theta\cos\theta\\
     0 & 0 & 0 & \sin\theta \cos\theta & \cos^2 \theta
     \end{pmatrix} \otimes \alpha\sum_{j} \sqrt{\frac{\hbar \Omega_j^2 Z_{0j}}{2}} (\hat{b}_j^\dagger + \hat{b}_j),
\end{equation}
\end{widetext}
where the $\hat{V}_j$ have been expressed in terms of the bosonic operators of the QBS [\eq{eq:LC_ladder_j}]. 
As in the spin-boson problem, the interactions in \eq{eq:simulator_hybrid} are diagonal in the $\{\ket{T_-}, \ket{S_0}\}$ simulation subspace and linear in the environmental degrees of freedom. However, thus far the physical model and the simulator are not identical. 

To overcome this, we propose to displace the QBS charge coordinates by applying a bias voltage 
\begin{equation}
    \Delta V = \Delta Q \sum_j \frac{1}{C_j}
\end{equation}
between the opposite ends of the QBS, which identically charges all the capacitors with $\Delta Q$. The new charge operator in the displaced coordinates
\begin{equation}\label{eq:charge_coord_shift}
    \hat{Q}_j' = \hat{Q}_j + \Delta Q
\end{equation}
still obey the desired commutation relations and  the displacement $\Delta Q$ can be conveniently chosen to yield either the spin-boson or the displaced harmonic oscillator. However, introducing $\Delta Q$ will result in $\hat{H}_\ti{QBS} \otimes \hat{\mathbf{1}}_\ti{QD} + \hat{H}_\ti{QD-QBS}$ static energy shifts that contribute to the $\epsilon_\ti{L/R}$ on top of the contributions due to the gate electrodes $\epsilon_\ti{L/R}^{g}$. Specifically,
\begin{subequations}
    \begin{equation}
    \epsilon_\ti{L} = \epsilon_\ti{L}^{g} + \sum_j \frac{(\Delta Q)^2 - 2 \alpha \Delta Q}{2 C_j},
    \end{equation}
    \begin{equation}
    \epsilon_\ti{R} = \epsilon_\ti{R}^{g} + \sum_j \frac{(\Delta Q)^2}{2 C_j},
    \end{equation}
\end{subequations}
and thus the detuning
\begin{equation}
    \epsilon_\ti{d} = \epsilon_\ti{d}^{g} + \sum_j \frac{\alpha \Delta Q}{C_j}.
\end{equation}
In these new coordinates $\hat{H}_\ti{QD}$ remains as defined in \eq{eq:simulator_bare}, and 
\begin{equation} \label{eq:HQBS_displaced}
    \hat{H}_\ti{QBS} = \sum_j \hbar \Omega_j \bigg( \hat{b}_j'^\dagger \hat{b}_j' + \frac{1}{2} \bigg),
\end{equation}
where the raising/lowering operators in the displaced charge coordinates are given by
\begin{subequations}\label{eq:b_prime}
\begin{equation}
    \hat{b}_j'^\dagger = \sqrt{\frac{1}{2\hbar Z_{0j}}}(Z_{0j} \hat{Q}_j' + i \hat{\Phi}_j),
\end{equation}
\begin{equation}
    \hat{b}_j' = \sqrt{\frac{1}{2\hbar Z_{0j}}}(Z_{0j} \hat{Q}_j' - i \hat{\Phi}_j).
\end{equation}
\end{subequations}
In turn, the resulting QD-QBS interaction term in
the \{$\ket{T_+}$, $\ket{T_0}$, $\ket{T_-}$, $\ket{S_0}$, $\ket{S_1}$\} basis is
\begin{widetext}
\begin{equation}\label{eq:QD_QBS_full}
    \hat{H}_\ti{QD-QBS} = \begin{pmatrix}
    \alpha - \Delta Q & 0 & 0 & 0 & 0 \\
    0 & \alpha - \Delta Q & 0 & 0 & 0 \\
    0 & 0 & \alpha - \Delta Q & 0 & 0 \\
    0 & 0 & 0 & \alpha \sin^2 \theta - \Delta Q & \alpha \sin \theta \cos \theta \\
    0 & 0 & 0 & \alpha \sin \theta \cos \theta & \alpha \cos^2 \theta - \Delta Q
    \end{pmatrix} \otimes \sum_{j} \sqrt{\frac{\hbar \Omega_j^2 Z_{0j}}{2}} ( \hat{b}_j'^\dagger + \hat{b}_j' ).
\end{equation}
\end{widetext}
By projecting into the simulation subspace, we obtain 
\begin{multline}\label{eq:QD_QBS_qs}
 \hat{H}_\ti{QD-QBS}^\ti{qs} 
    = [ (\alpha - \Delta Q) \ket{T_-} \bra{T_-} \\
    + (\alpha \sin^2 \theta - \Delta Q) \ket{S_0} \bra{S_0} ] \otimes \sum_{j} \sqrt{\frac{\hbar \Omega_j^2 Z_{0j}}{2}} (\hat{b}_j'^\dagger + \hat{b}_j').
\end{multline}
In this charged coordinate system, one can manipulate $\Delta Q$ to yield different types of interactions. When $\Delta Q=\alpha$ the interaction is identical to the displaced harmonic oscillator. In turn, for $\Delta Q= \half \alpha (1 + \sin^2 \theta) $ the interaction is identical to the spin-boson problem. By separating $\hat{B}_\ti{qs}$ [\eq{eq:TCF_LC}] from \eq{eq:QD_QBS_qs}, we can now also specify the undetermined constant in \eq{eq:TCF_LC} as $\kappa = \alpha \cos^2 \theta$ for the displaced harmonic oscillator and $\kappa = \half \alpha \cos^2 \theta$ for the spin-boson problem.

\subsection{Choosing simulator parameters}\label{subsection:choose_param}

The total Hamiltonian of the simulator is
\begin{multline}\label{eq:Hsimulator}
    \hat{H}_\ti{qs} = \hat{P}^\dagger \hat{H}_\ti{exp} \hat{P} = \\ \hat{H}_\ti{QD}^\ti{qs} \otimes \hat{\mathbf{1}}_\ti{QBS} + \hat{\mathbf{1}}_\ti{QD}^\ti{qs} \otimes \hat{H}_\ti{QBS} + \hat{H}_\ti{QD-QBS}^\ti{qs},
\end{multline}
where $\hat{P}$ is the projection operator onto the simulation subspace. The simulator Hamiltonian is specified by Eqs.~\eqref{eq:2state_qs}, \eqref{eq:HQBS_displaced}, \eqref{eq:QD_QBS_qs}, and \eqref{eq:Hsimulator}.
Thus, as desired, if we associate $\ket{g} \rightarrow \ket{T_-}$ and $\ket{e} \rightarrow \ket{S_0}$ we can exactly map $\hat{H}$ in \eq{eq:H_full} to $\hat{H}_\ti{qs}$ in \eq{eq:Hsimulator}. An important feature of the proposed mapping is that it enables to continuously tune Hamiltonian parameters through experimentally accessible variables. We now discuss how to choose experimental variables needed to reproduce a given target Hamiltonian.

\begin{table*}[htbp]
\caption{ \label{tab:table1} 
Typical ranges for physical parameters in energy/charge transport processes for photosynthetic complexes at $T$ = 300 K, and how they translate in the quantum simulator operating at cryogenic temperature $T_\ti{qs}$ = 60 mK. The quantity $\gamma = T/T_\ti{qs} = 5 \times 10^3$ denotes the ratio between temperatures of the target problem and the quantum simulator. Tuning the temperature ratio can be used to map a wide range of target  parameters.
}
\begin{ruledtabular}
\begin{tabular}{lcccccc}
&\multicolumn{2}{c}{Target system}& &\multicolumn{2}{c}{Quantum simulator} \\
Element & Maximum & Resolution &  Scaling & Maximum & Resolution \\ \hline
Simulation time & 10 ps & 2 fs & $\gamma$ & 50 ns & 10 ps \\ \hline
Molecular energy span ($\Delta$) & \makecell{125 meV \\ 1000 cm$^{-1}$} & \makecell{125 $\mu$eV \\ 1 cm$^{-1}$} & $1/\gamma$ & \makecell{25 $\mu$eV \\ 6 GHz} & \makecell{2.5 neV \\ 6 MHz} \\ \hline
Molecular coupling ($\eta$)& \makecell{25 meV \\ 200 cm$^{-1}$} & \makecell{125 $\mu$eV \\ 1 cm$^{-1}$} & $1/\gamma$ & \makecell{5 $\mu$eV \\ 1.2 GHz} & \makecell{2.5 neV \\ 6 MHz} \\ \hline
Bath frequency & \makecell {250 meV \\ 2000 cm$^{-1}$} & \makecell{125 $\mu$eV \\ 1 cm$^{-1}$} & $1/\gamma$ & \makecell{50 $\mu$eV \\ 12 GHz} & \makecell{2.5 neV \\ 6 MHz} \\ \hline
Reorganization Energy & \makecell {100 meV \\ 800 cm$^{-1}$} & \makecell{125 $\mu$eV \\ 1 cm$^{-1}$} & $1/\gamma$ & \makecell{20 $\mu$eV \\ 4.8 GHz} & \makecell{1 neV \\ 2.4 MHz}
\end{tabular}
\end{ruledtabular}
\end{table*}

At this stage, it is useful to introduce the sensitivity of the quantum simulator to the fluctuations of the QD energy as an important target feature of the simulator. This is important because such sensitivity determines how the simulator couples to its natural environment and to the QBS, and will guide the physical requirements in QBS design. We define the sensitivity as how the $S_0$--$T_-$ energy gap $\Delta_\ti{qs}$ changes with the detuning $\epsilon_\ti{d}$, i.e.
\begin{equation}\label{eq:sensitivity}
    k_\ti{s} \equiv \bigg| \frac{d \Delta_\ti{qs}}{d \epsilon_\ti{d}} \bigg| = \frac{1}{2}\bigg(1 + \frac{\epsilon_\ti{d}}{\sqrt{\epsilon_\ti{d}^2 + 4 t_\ti{c}^2}}\bigg) = \cos^2 \theta
\end{equation}
where the latter equality follows by combining \eq{eq:sensitivity} with \eq{eq:mixingangle}.
The sensitivity can be tuned in the QD setup by changing $\epsilon_\ti{d}$. Increasing the sensitivity $0 < k_\ti{s} < 1$ has the desirable consequence of reducing the physical size of the QBS required to achieve a given physical effect. However, increasing $k_\ti{s}$ also has the undesirable effect of reducing the coherence time of the simulator states as it makes the influence of the uncontrollable electrical noise in the experimental setup more important. Optimizing $k_\ti{s}$ is essential for the development of a useful simulator.

By expressing the mixing angle $\theta$ in terms of the sensitivity ($\sin \theta = \sqrt{1 - k_\ti{s}}$, $\cos \theta = \sqrt{k_\ti{s}}$), we find that the external control fields needed to realize a given set of $\Delta$, $\eta$, and $k_\ti{s}$, with a temperature ratio $\gamma$ and tunnel coupling  $t_\ti{c}$ are
\begin{equation}\label{eq:map_sys_param}
    \begin{gathered}
    \epsilon_\ti{d} = \frac{t_\ti{c}(2k_\ti{s} - 1)}{\sqrt{k_\ti{s}(1 - k_\ti{s})}}, \\ 
    B_\ti{avg} = \frac{1}{g \mu_B} \bigg( t_\ti{c} \sqrt{\frac{k_\ti{s}}{1 - k_\ti{s}}} + \frac{\Delta}{\gamma} \bigg), \\
    \Delta B = \frac{\eta}{g \mu_B \gamma}\sqrt{\frac{8}{1 - k_\ti{s}}}.
    \end{gathered}
\end{equation} 

\Eqs{eq:spd_vs_RLC}{eq:map_sys_param} with $\kappa = \alpha k_\ti{s}$ for the displaced harmonic oscillator or $\kappa = \half \alpha k_\ti{s}$ for the spin-boson problem provide a complete mapping between the original problem [\eq{eq:H_full}] and the quantum simulator [\eq{eq:2state_qs}].

\section{Emulating the Simulator}
\label{stn:emulations}

\subsection{Simulation and Physical Timescales and Energies}

To illustrate how physical timescales and energies are mapped into the quantum simulation setup,  consider Table~\ref{tab:table1} detailing the physical requirements of the map. For illustration purposes, we focus on target parameters needed to capture excitonic dynamics in photosynthetic complexes operating at $T=300$ K in a simulator operating at $T_\text{qs}=60$ mK. To establish a map between chemical environments and our QBS, energies and timescales need to be scaled according to the temperature ratio $\gamma= T/T_\text{qs}$. The dynamics in the simulator will be $\gamma$ times slower, and the characteristic energies $1/\gamma$ smaller than in the physical system. Importantly, the characteristic timescales and energies of such a molecular system map into timescales and energies that are accessible in existing QD setups~\cite{Kloeffel2013} making it a viable simulation strategy. By tuning the temperature ratio $\gamma$ it is possible to map a wide range of target parameters into an experimentally accessible operation conditions for the QDs. 


\subsection{Quantum Dynamics of the Simulator}\label{subsection:emulation}

To understand the practical operation of the simulator for different conditions and the extent to which the quantum dynamics can be confined to the desired simulation subspace, we now emulate the full five-state QD Hamiltonian interacting with the QBS [\eq{eq:H_QD_QBS_hybrid}] and compare the dynamics to that of the target system. 
While emulations in the presence of highly structured environments are beyond the reach of state-of-the-art computational methods,
for simple bath spectral densities such emulations can be conducted using the hierarchical equations of motion (HEOM).\cite{Tanimura2020} In these HEOM simulations it is necessary to go beyond usual high-temperature approaches and include all low-temperature corrections. This is needed because the large energy separation between the spectator states and the $S_0/T_-$ manifold prevents invoking the high temperature approximation.

For definitiveness, we study the relaxation to thermal equilibrium of a molecule initially in the excited state. Specifically, we choose the displaced Harmonic oscillator model [Eqs.~(\ref{eq:H_full})--(\ref{eq:BSD}) with $\hat{S} = \ket{e}\bra{e}$] with system parameters set as $\Delta = 10$ meV and $\eta = 5$ meV. The system-bath coupling is taken to be described by a Drude-Lorentz spectral density
\begin{equation}\label{eq:DL_BSD}
    J(\omega) = \frac{2 \lambda}{\pi} \frac{\omega_\ti{c} \omega}{\omega^2 + \omega_\ti{c}^2},
\end{equation}
with reorganization and cutoff energies $\lambda = 5$ meV and $\hbar\omega_\ti{c} = 10$ meV, respectively. These parameters are similar to those extracted from typical natural photosynthetic complexes.\cite{Adolphs2006,Kell2013} The target (300 K) and simulator (60 mK) temperatures are those considered in Table \ref{tab:table1}.
All HEOM calculations were conducted with a hierarchy depth of 10 and the low-temperature correction by using 6$^\ti{th}$ order Pad{\'e} expansion of the Bose-Einstein distribution function.\cite{Hu2010} For all simulations, the dynamics was propagated until the physical time of $1.5$ ps, which corresponds to a  $7.5$ ns simulation time. Such time scale is well within the usual regime of coherent operation and resolution of QDs in experiments.

To monitor the  dynamics we follow the population inversion $\langle \hat{\sigma}_z (t)\rangle$ of the two-level molecule as it relaxes to thermal equilibrium. To quantify the accuracy of the quantum simulation, we compute the time-dependent fidelity 
\begin{equation}\label{eq:fidelity}
    F(t) = \bigg( \ti{Tr} \sqrt{ \hat{\rho}^{1/2}_\ti{s}(t) \hat{\rho}^\ti{qs}_\ti{s}(t) \hat{\rho}^{1/2}_\ti{s}(t) } \bigg)^2,
\end{equation}
between the HEOM reduced density matrix for the system $\hat{\rho}_\ti{s}$ and that of the simulator $\hat{\rho}^\ti{qs}_\ti{s}(t)$ in the simulation subspace. The fidelity $0\le F(t) \le 1$ increases as the simulations becomes more accurate. To quantify the ability of the simulator to contain the dynamics in the simulation subspace, we further quantify the total leakage of population into the spectator states.

\begin{figure}[htbp]
\includegraphics[width=\columnwidth]{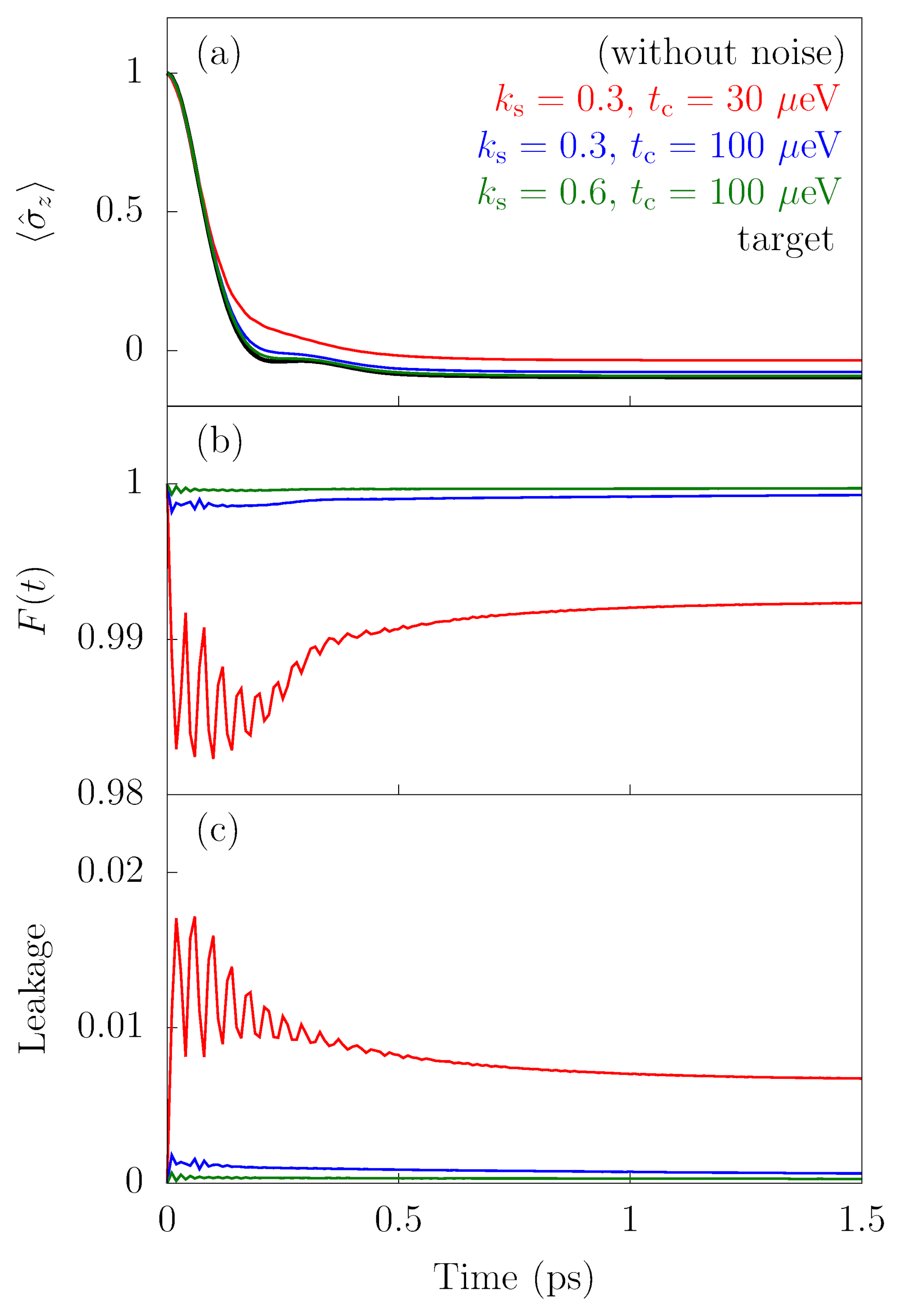}
\caption{Performance of the quantum simulator for three different sensitivities $k_\ti{s}$ and tunnel couplings $t_\ti{c}$. (a) Population inversion $\langle \hat{\sigma}_z (t)\rangle$ in the target dynamics (black line) and the simulator (colored lines); (b) Fidelity $F(t)$ and; (c) Total population leaked out of the simulation subspace. Time is physical time. Simulation time is obtained by scaling by $\gamma=5 \times 10^3$. In \fig{fig:2d_all}a, the three specific choices of $t_\ti{c}$ and $k_\ti{s}$ used are marked by the colored (red, blue and green) dots.}
\label{fig:emul_pristine_t}
\end{figure}

In constructing the simulator, there is freedom in the choice of tunnel coupling $t_\ti{c}$ and sensitivity $k_\ti{s}$. However,  the choice influences simulation accuracy and determines the experimental requirements to build the simulator (see \stn{stn:experimentalneeds}).
\Fig{fig:emul_pristine_t} shows the population inversion, fidelity and total leakage during the dynamics for three select choices for $t_\ti{c}$ and  $k_\ti{s}$. As shown, the simulator is able to accurately capture the target population dynamics (black line) in the three  different regimes of operation selected, with fidelities larger than 0.98 in all cases. The disagreement between the simulator and exact dynamics is most apparent for relatively low $t_\ti{c}=30$ $\mu$eV (red lines). We observe that a decrease in the accuracy of the simulation is correlated to population leaking out of the simulation space (\fig{fig:emul_pristine_t}c). By contrast, for $t_\ti{c} = 100$ $\mu$eV the simulation accuracy is high for both $k_\ti{s} = 0.3$ (blue lines) and $k_\ti{s} = 0.6$ (green lines). 


\begin{figure*}[htbp]
    \includegraphics{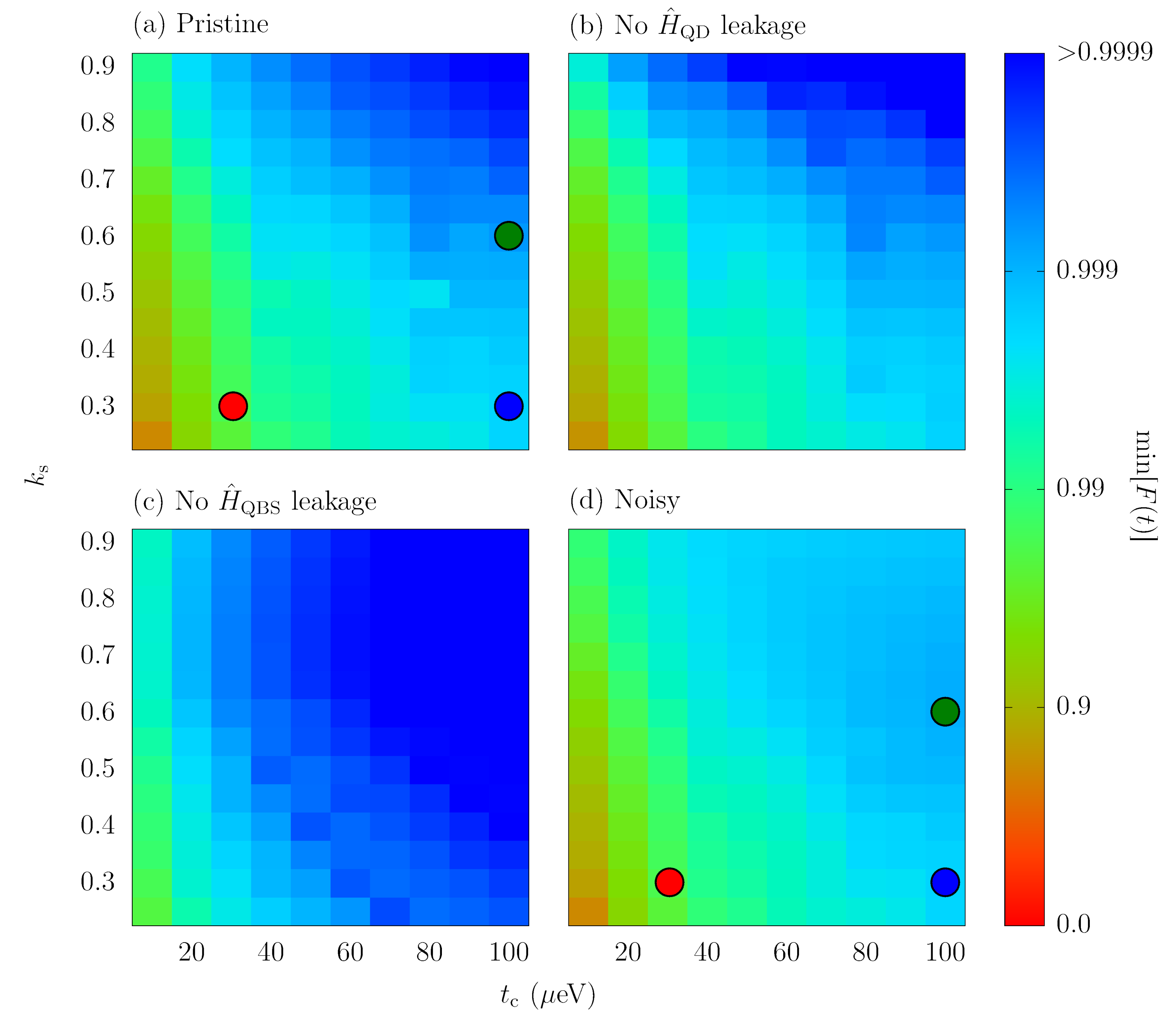}
        \caption{Accuracy of the simulator for different  sensitivities $k_\ti{s}$ and tunnel couplings $t_\ti{c}$. The plots show heat maps of the minimum fidelity [\eq{eq:fidelity}] throughout the simulation period  $\ti{min}[F(t)]$ for simulators (a) in pristine form and (b-c) without coupling to spectator states  ($\ket{S_1}, \ket{T_+}, \ket{T_0}$) through either (b) the QD Hamiltonian $\hat{H}_\ti{QD}$ [\eq{eq:simulator_bare}] or (c) the QD-QBS interaction $\hat{H}_\ti{QD-QBS}$ [\eq{eq:simulator_hybrid}]. Panel (d) quantifies simulation accuracy in the presence of electrical noise $\sigma_\epsilon = 2$ $\mu$eV added to the detuning $\epsilon_\ti{d}$. The color map is based on a logarithmic scale.}
\label{fig:2d_all}
\end{figure*}

To characterize how the choice of tunnel coupling $t_\ti{c}$ and sensitivity $k_\ti{s}$ influences simulation accuracy, we performed extensive emulations for $0.25 \le k_\ti{s} \le 0.9$ and $10$ $\mu$eV $\le t_\ti{c} \le$ $100$ $\mu$eV. Simulation accuracy was quantified by extracting  the minimum fidelity  $\ti{min}[F(t)]$ throughout the simulation time. We stress that this is an over-estimate of infidelity, since it is an extreme value instead of an average.
\Fig{fig:2d_all}a shows the heat map of $\ti{min}[F(t)]$ as a function of $k_\ti{s}$ and $t_\ti{c}$. The three specific choices of $t_\ti{c}$ and $k_\ti{s}$ in \fig{fig:emul_pristine_t} are marked by the colored (red, blue and green) dots in \fig{fig:2d_all}a.  We observe that the accuracy of simulation increases with both $k_\ti{s}$ and $t_\ti{c}$, and even exceeds fidelities of 99.9\% in the upper-right part of the map demonstrating high simulation accuracy in this regime of operation.

As shown in \fig{fig:emul_pristine_t}b-c, the decay of $F(t)$ is strongly correlated with the leakage to the spectator states.  Suppressing this leakage is therefore the most crucial task for improving the quality of the simulation. We can identify two sources of leakage: (i) the magnetic couplings in $\hat{H}_\ti{QD}$ involving the spectator states [\eq{eq:simulator_bare}] and (ii) the residual interaction between $\ket{S_0}$ and $\ket{S_1}$ in $\hat{H}_\ti{QD-QBS}$  [\eq{eq:QD_QBS_full}]. To understand their relative contributions, we repeated the emulations eliminating either element from the simulator Hamiltonian. \Fig{fig:2d_all}b and c shows results without leakage due to $\hat{H}_\ti{QD}$ and $\hat{H}_\ti{QD-QBS}$, respectively. As seen, leakage due to $\hat{H}_\ti{QD}$ has a minor effect in the simulation (\fig{fig:2d_all}b). By contrast, leakage due to  $\hat{H}_\ti{QD-QBS}$ is the dominant effect as removing this effect  greatly expands the ($t_\ti{c}$, $k_\ti{s}$) region of high simulation fidelity (\fig{fig:2d_all}c).

\subsection{Effect of uncontrollable noise}

In practical realizations, the simulator will also suffer from the inherent uncontrollable noise of the experimental setup. It is thus important to determine if the simulation strategy remains viable in the presence of the decoherence due to a noisy environment.

For QD systems, the main source of decoherence is the electrical noise that causes fluctuations of $\epsilon_\ti{d}$ and, in turn, $\Delta_\ti{qs}$. A recent noise spectroscopy study\cite{Connors2022} has demonstrated that this noise mainly arises from low-frequency noise ($\ll 10^6$ Hz), whose time scale is much slower compared to those relevant to the simulation (Table~\ref{tab:table1}). Based on this observation, we model how the noise affects the quality of the simulation by perturbing $\epsilon_\ti{d}$ with static noise $\delta \epsilon_\ti{d}$ which follows a Gaussian distribution of zero mean and standard deviation of $\sigma_\epsilon = 2$ $\mu$eV\cite{Jock2018}. Practically, this was achieved by discretizing the Gaussian distribution into 10 points uniformly spaced by 0.5 $\mu$eV, and conducting the emulation at each point with the detuning of $\epsilon_\ti{d} + \delta \epsilon_\ti{d}$ while not altering $t_\ti{c}$, $B_\ti{avg}$, and $\Delta B$. The system density matrix with noise was calculated as the weighted average $\hat{\rho}_\ti{s}(t) = \sum_{n=1}^{10} w_n \hat{\rho}_{\ti{s}n}(t)$ where $\hat{\rho}_{\ti{s}n}(t)$ denotes the system density matrix obtained from the emulation at the $n$-th point, and $w_n$ is the appropriate weight for this point determined by integrating the Gaussian distribution.

\begin{figure}[htbp]
\includegraphics[width=\columnwidth]{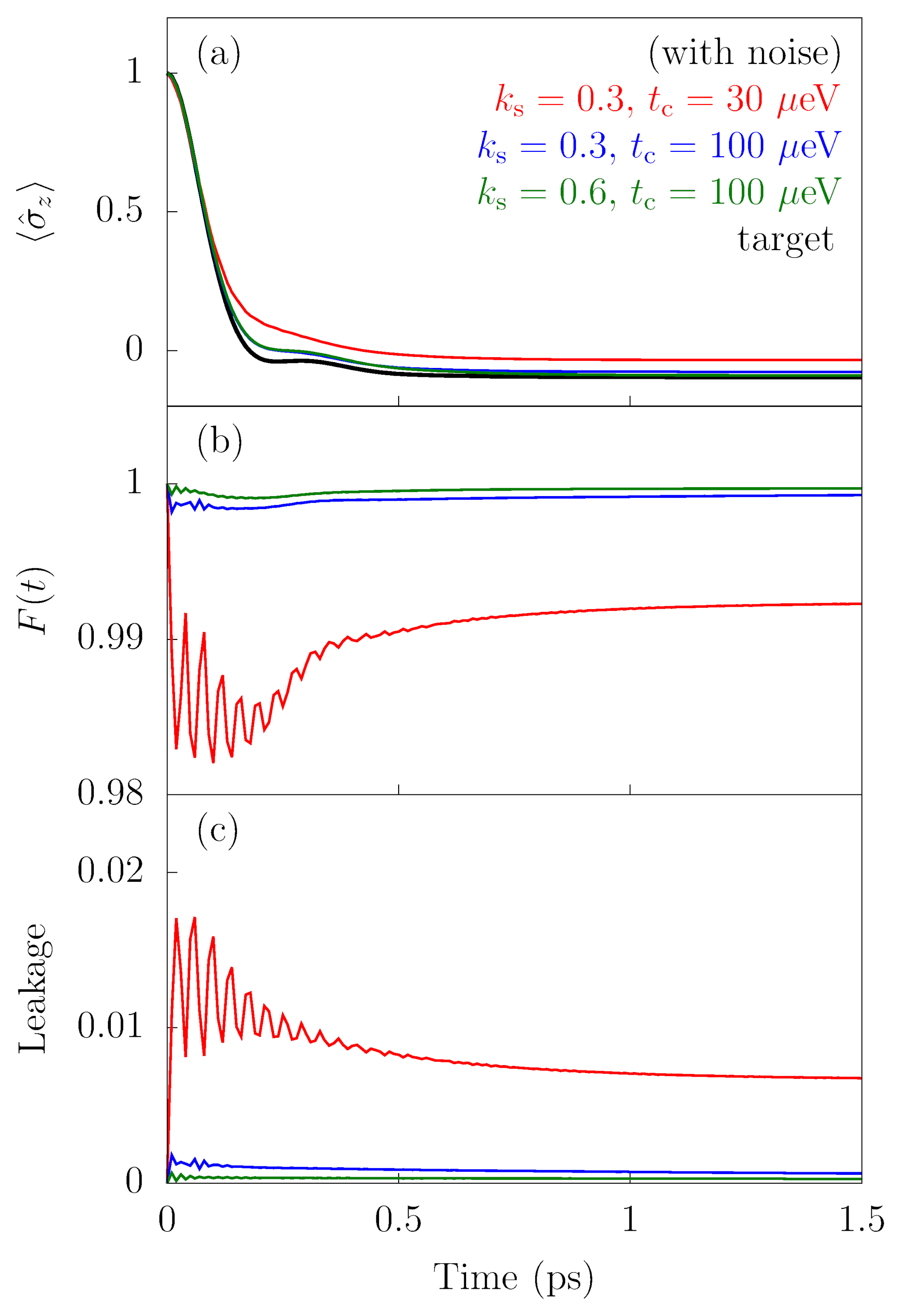}
\caption{\label{fig:emul_noise_t} Performance of the quantum simulator in the presence of electrical noise that leads to fluctuations of the detuning $\epsilon_\ti{d}$ with a standard deviation $\sigma_\epsilon = 2$ $\mu$eV. The panel distribution is identical to that in \fig{fig:emul_pristine_t}. In \fig{fig:2d_all}d, the three specific choices of $t_\ti{c}$ and $k_\ti{s}$ used are signaled by the colored (red, blue and green) dots.}
\end{figure}


\Fig{fig:emul_noise_t} shows the emulated dynamics of the simulator in the presence of noise, under conditions otherwise identical to those in \fig{fig:emul_pristine_t}. In general, the influence of electrical noise leads to only modest changes in the dynamics. The effect of the noise is the most important for the case with $k_\ti{s} = 0.6$ and $t_\ti{c} = 100$ $\mu$eV (green curve), which showed almost perfect agreement with the target dynamics in \fig{fig:emul_pristine_t}a but now exhibits slight deviation at early times in \fig{fig:emul_noise_t}a. The effect of noise is not as prevalent in the other two simulation conditions (red and blue curves) which already exhibit some amount of leakage-induced error. Figures~\ref{fig:emul_noise_t}b and \ref{fig:emul_noise_t}c show that the noise has a negligible effect on the amount of leakage.

To determine the conditions optimal for the experimental operation of the simulator, we computed the minimum fidelity min$[F(t)]$ in the presence of noise for varying $k_\ti{s}$ and $t_\ti{c}$. \Fig{fig:2d_all}d shows the resulting heat map for min$[F(t)]$. A comparison with \fig{fig:2d_all}a shows that, overall, the noise slightly deteriorates the quality of the simulation. The green curve in \fig{fig:emul_noise_t} is representative of the best simulator performance in our emulations. The effect of the noise is the most prominent for large sensitivities, as this is the region where the quantum dot energy levels are the most susceptible to fluctuations in the detuning. Remarkably, even in the presence of noise we observe that min[$F(t)$] still exceeds 99\% for most of the simulation conditions and even 99.9\% for $0.55 \le k_\ti{s} \le 0.7$ with $t_\ti{c} = 100$ $\mu$eV.



These emulation results illustrate the utility of the simulator to model condensed phase dynamics. As shown, even in the presence of noise, it is possible to choose sensitivities and tunnel couplings in which the leakage into non-simulator states is suppressed to $<0.5$\% (\fig{fig:emul_noise_t}c), leading to small errors in the simulated dynamics (\fig{fig:emul_noise_t}b). The data shows that a larger $t_\ti{c}$ lead to better agreement between the target problem and the quantum simulator, as increasing $t_\ti{c}$ leads to a larger energy separation between $S_0$ and $S_1$ and suppresses leakage into non-simulator states. In turn, there exists an optimal value of $k_\ti{s}$ for a fixed $t_\ti{c}$. This is because larger $k_\ti{s}$ suppresses the leakage by decreasing the  strength of residual coupling between $S_0$ and $S_1$ due to $\hat{H}_\ti{QD-QBS}$, but also increases the amount of electrical noise felt by the simulation subspace.

\section{Experimental Considerations}
\label{stn:experimentalneeds}

To mimic the system, the simulator requires a coherent DQD system in the presence of an homogeneous magnetic field and a magnetic field gradient along the current path. These QDs must be electrically connected to an array of quantum circuits of high impedance and with varying frequencies. The electrical noise in the QDs must be small enough such that the effect of unwanted decoherence remains negligible throughout the simulation time. In addition, the simulator components and their interactions in the experiments must closely follow the model Hamiltonian [Eqs.~(\ref{eq:HQBS_displaced}) and (\ref{eq:QD_QBS_full})].

In designing the simulator one has freedom on the sensitivity ($k_\ti{s}$), temperature scaling ($\gamma$) and tunnel coupling ($t_\ti{c}$) employed. However, the choice influences the quality of the simulation (see \stn{stn:emulations}) and the experimental requirements to build it, including the required QBS size, magnetic fields and coherence times. The choice of $k_\ti{s}$, $\gamma$ and $t_\ti{c}$ ideally minimizes experimental requirements and maximizes the accuracy of the simulation. 

We now discuss how varying these parameters imposes requirements on experimental coherence, magnetic fields, and QBS size as well as the consequences on simulation accuracy. We also discuss the effect of possible non-ideal behavior of the simulator components and possible ways to mitigate it if significant.

\subsection{Coherence times}
The simulation time is affected by how long the simulator can preserve its quantum coherence in the presence of noise. The charge noise is the dominant source of decoherence with time scales $\sim 10$s of ns. By contrast, decoherence effects arising from hyperfine fluctuations of residual nuclear spins is much slower $\sim 100$s of ns or longer\cite{Maune2012}. We thus focus on the decoherence due to charge noise.

The standard deviation of $\epsilon_\ti{d}$ due to electrical noise for state-of-the-art QD devices is $\sigma_\epsilon \sim 2$ $\mu$eV,\cite{Higginbotham2014,Jock2018} and has a weak temperature dependence for $T < 1$ K.\cite{Petit2020} The effective magnitude of fluctuation felt by the simulation subspace is $\sigma = k_\ti{s} \sigma_\epsilon$, which leads to decoherence time $\tau_\ti{d} \sim \frac{2 \pi \hbar}{\sigma}$ of the simulator. Therefore, an estimate of the coherence time of the target system that the simulator can support is
\begin{equation}\label{eq:t_coh}
    \tau \sim \frac{\tau_\ti{d}}{\gamma} = \frac{2 \pi \hbar}{k_\ti{s} \sigma_\epsilon \gamma},
\end{equation}
which illustrates that the simulation time can be enhanced by decreasing $k_\ti{s} \in [0,1]$ as it slows down the decoherence or decreasing $\gamma$ as it speeds up laboratory simulation time. 

For example, for a sensitivity $k_\ti{s} = 0.6$, the decoherence time is $\tau_\ti{d} \sim 3.4$ ns. The highest operation temperature for the QDs is 1 K,\cite{Yang2020} which yields a $\gamma = 3 \times 10^2$ for a simulation of a room-temperature process. Thus, in this case the maximum simulation time is $\sim 11$ ps suggesting that the simulator will be accurate in capturing both the early times and asymptotic behavior of the dynamics of microscopic open quantum systems. Even when the QDs are operated at the usual temperature of 100 mK, the maximum simulation time is 1.1 ps that is more than suitable to model most molecular processes.



\subsection{Magnetic fields}

A homogeneous magnetic field is employed in the simulator to modulate the energy gap and separate simulation states from spectator states. In turn, a magnetic field gradient is needed to introduce couplings between the simulator states. Using \eq{eq:map_sys_param} one can develop a quantitative estimate for the experimental requirements of the simulator. For definitiveness, consider the example in \fig{fig:2d_all} with $t_\ti{c} = 100$ $\mu$eV and $k_\ti{s}=$ 0.6. For Si- (or GaAs-) based QDs, the simulator requires $B_\ti{avg} =$ 1.08 T (or 5.00 T) and $\delta B =$ 38.6 mT (or 179 mT). The less demanding requirements in silicon arise because the Land{\'e} $g$-factor is larger ($g=2.00$)\cite{Wilson1961} with respect to that of GaAs  ($g =$ 0.43).\cite{Yugova2007} $B_\ti{avg}$ stronger than 5 T can be routinely generated in experiments and are not expected to be a limiting factor in building the simulator. By contrast, state-of-the-art magnetic field gradients are limited to 1 mT/nm and they impose practical limits on $\eta$ that can be modeled. For example, the usual distance between Si QDs is $\sim 100$ nm, which implies that the maximum magnetic field change between consecutive QDs is $\sim 100$ mT. Since $k_\ti{s} > 0$ it follows from \eq{eq:map_sys_param} that there is an upper limit for $\eta$ given a $\Delta B$,
\be
    |\eta| < \frac{\gamma g \mu_B |\Delta B|}{2 \sqrt{2}}.
\ee
For a limiting $\gamma$ values of 3$\times10^2$--3$\times10^4$, the upper limit of $\eta$ can be in the range of 1.2--120 meV. The larger the $\gamma$ the larger the $\eta$ that can be modeled.

\subsection{QBS Requirements}\label{subsection:QBS_req}

The proposed QBS requires arrays of serially connected resonators that generate the desired level of voltage fluctuation. To understand the frequency and impedance requirements to build the QBS, it is useful to consider the  parameters needed in molecular modeling. For chemical problems, the  bath frequencies are in the $10^{-4} < \omega_j< 0.4$ eV range or $\frac{10^{-4}}{\gamma}< \Omega_j< \frac{0.4}{\gamma}$ eV for the QBS. The frequency of operation of the materials needed for the QBS will impose restrictions on the choice of $\gamma$. For example, high impedance resonators constructed with superinductors (e.g. NbTiN~\cite{Samkharadze2016} and granular aluminum~\cite{Gruenhaupt2019}) have frequencies up to 20 GHz, which implies $300 < \gamma < 5 \times 10^3$, where the lower limit is imposed by the magnetic field requirements and the upper one by the frequencies of the superinductor resonators.

We now estimate the number of resonators needed per color to model complex chemical environments. The strength of the coupling between electrons and environmental modes is quantified by the Huang-Rhys factors $s_j \equiv (g_j / \omega_j)^2$. The  characteristic impedance required in the QBS to achieve these Huang-Rhys factors at a given frequency is [Eqs.~\eqref{eq:BSD}, \eqref{eq:TCF_0} and \eqref{eq:TCF_LC}]
\begin{equation}\label{eq:Z_in_HR}
    Z_{j0} = 2 \hbar \bigg( \frac{n g_j}{\alpha k_\ti{s} \gamma \Omega_j} \bigg)^2 = 2 \hbar s_j \bigg( \frac{n}{\alpha k_\ti{s}} \bigg)^2,
\end{equation}
where $n = 1$ for the displaced harmonic oscillator and $n = 2$ for the spin-boson model.
The required $Z_{j0}$ increases with the Huang-Rhys factor $s_j$, decreases by choosing a larger sensitivity, and is independent of the temperature scaling $\gamma$. 

To proceed, it is useful to decompose the spectral density $J(\omega) = J_\ti{L}(\omega) + J_\ti{H}(\omega)$ into a component due to sharply peaked high frequency modes such as intramolecular vibrations $J_\ti{H}(\omega)$, and an unstructured broad low frequency component due to solvent $ J_\ti{L}(\omega)$. In one partition,\cite{MontoyaCastillo2015} $J_\ti{L}(\omega) = S(\omega, \omega^*) J(\omega)$ and $J_\ti{H}(\omega) = [1 - S(\omega, \omega^*)] J(\omega)$ where the splitting function $S(\omega, \omega^*)= [1 - (\omega / \omega^*)^2]^2$ for $\omega<\omega^\star$ and zero otherwise.

For high frequency modes of typical molecules $0.001 < s_j < $ 0.1 which requires a characteristic impedance of $2 \times 10^3 < Z_{j0} < 2 \times 10^5$ Ohms (assuming the realistic values $\alpha = 0.1$ eV/V, $k_\ti{s}=$ 0.6 and $n = 1$). For state-of-the-art resonators,\cite{Samkharadze2016,Kamenov2020,Peruzzo2020} $Z_0$ ranges from $10^3$ to $10^4$, and therefore up to $10^2$ serially connected resonators are needed to construct the QBS per color. Given sufficiently high $Z_{j0}$, the contribution of each color to the decoherence can be modulated by scaling down the signal by attaching a tunable voltage divider circuit.

By contrast, for low frequency components the Huang-Rhys factors can easily exceed $s_j \gg 1$ making the above approach intractable. To overcome this, we take advantage of the fact that for these modes $\hbar \Omega \ll k_\ti{B} T_\ti{qs}$, spontaneous emission processes are negligible and classical noise approximations useful.\cite{Gu2019} Specifically, we propose to model this component of the spectral density by introducing classical noise $V(t)$ to the voltage of the gate electrode that controls the dot energy. This assumes that the low-frequency component does not lead to sizable spontaneous emission of phonons or, equivalently, that the voltage time correlation function is real and satisfies [\eq{eq:TCF_0}]
\begin{multline}\label{eq:J_lowfreq_tcf}
    \langle V(t) V(0) \rangle = \frac{1}{\kappa^2 \gamma^2} \ti{Re} \big[ \langle \hat{B}(t / \gamma)\hat{B}(0) \rangle_{T} \big] \\
    = \frac{\hbar}{\gamma} \bigg( \frac{n}{\alpha k_\ti{s}} \bigg)^2 \int_0^{\infty} J_\ti{L}(\gamma \Omega) \coth \bigg( \frac{\hbar \Omega}{2 k_\ti{B} T_\ti{qs}} \bigg) \cos \Omega t \: d\Omega.
\end{multline}

\subsection{Tunnel Coupling}

For simulation purposes, increasing the tunnel coupling $t_\ti{c}$ is desirable since it improves the quality of the simulation by broadening the energy separation between $S_0$ and $S_1$. However, for measurement purposes smaller values of $t_\ti{c}$ are preferred as they are required for 
initializations of the QDs and the rapid readouts of the state populations by taking advantage of Pauli spin blockade\cite{Ono2002} in the QDs. Currently, the range of $t_\ti{c}$ that  allows useful measurement/initialization is 10--20 $\mu$eV.\cite{Borselli2011,Mills2019a} By contrast, larger $t_\ti{c}$ values enhance the simulation fidelity and add more flexibility in choosing $k_\ti{s}$.
Measurements for larger $t_\ti{c}$ can be accomplished using read-out based on spin-selective tunneling~\cite{Elzerman2004}. Alternatively, different $t_\ti{c}$ for simulation and measurement can be employed.

\subsection{Sensitivity}

Choosing an optimal value of $k_\ti{s}$ must balance leakage, simulator accuracy and hardware requirements. Specifically, increasing the sensitivity $k_\ti{s}$ reduces leakage and also the minimum required size for the QBS which are both desirable aspects in simulator design. However, it also reduces the simulation time the setup can support [\eq{eq:t_coh}] and increases the $B_\ti{avg}$ and, more importantly, the $\Delta B$ needed to realize desired values of $\Delta$ and $\eta$ [\eq{eq:map_sys_param}]. For example, emulations in \stn{stn:emulations} have shown that sensitivities in the range of $0.55 \le k_\ti{s} \le 0.7$ exhibit good simulator performance with large $t_\ti{c}$ while requiring moderate QBS sizes to achieve a given level of desired decoherence. However, they also have the downside of requiring strong magnetic fields to realize a given $\hat{H}_\ti{s}$, especially for GaAs-based QDs with a small $g$-factor. Using Si-based QDs and increasing the $t_\ti{c}$ will broaden the range of the sensitivity that can be used in practical simulations.

\subsection{Gate Voltages}\label{subsection:gatevoltage}
In the experiment, the effect of changing the voltage of a gate electrode (\fig{fig:QD}b) often stretches across multiple QDs. Nevertheless, energies of individual QDs can still be independently tuned by finding appropriate linear combinations of the gate voltages, which are called virtual gates.\cite{Mills2019a} Similarly, while connecting the QBS to a single electrode would not exclusively affect a single QD as assumed in \eq{eq:qb_qds}, this type of coupling can still be realized  by  connecting the QBS to multiple gate electrodes and scaling the coupling according to the ratios between the linear combination coefficients of the virtual gates.

\subsection{Real Circuits}
Until now, we have treated the $LC$ circuits as perfect harmonic oscillators [\eq{eq:H_LC}] with fully controllable circuit parameters. However, experimentally fabricated circuits can deviate from such an ideal model. We now discuss such effects on the effectiveness of the simulator.

\paragraph{Resonator anharmonicity.} We expect possible anharmonicities in the QBS circuits to play a negligible role in the simulator. Deviations from the harmonic model for the QBS lead to uneven spacing between the phononic quantum states. For our QBS we estimate the anharmonicity $\big| \frac{\omega_2 - 2 \omega_1}{\omega_1} \big| < 10^{-3}$ \cite{Sage2011} where $\hbar\omega_n$ is the energy of the $n$-th quantum state. Assuming that the energy difference between the consecutive states decreases at a uniform rate of $10^{-3}\hbar \omega_1$, the effect of anharmonicity will not become prevalent unless $\sim 10^2$ or more quantum states are significantly populated. Therefore, to a good approximation the $RLC$ circuits behave as ideal harmonic oscillators except for the ones with very small energy spacing compared to the thermal energy ($\hbar \Omega \ll k_\ti{B} T_\ti{qs}$). Even for such cases, the anharmonicity will have a negligible effect as the characteristic time scale of the resonators would be much slower than that of the dynamics. At this limit, the oscillators only exert quasistatic noise and do not readily exchange energy with the system.

\paragraph{Inherent dissipation} of the $LC$ resonators broadens their impedance $\ti{Re}[Z(\Omega)]$ from perfect delta functions [\eq{eq:LC_impedance}]. This dissipation is measured by the quality factor (the ratio between the characteristic frequency and bandwidth) which  is $q= 10^4 - 10^5$ for state-of-the art resonators\cite{Samkharadze2016,Gruenhaupt2019,Peruzzo2020}. By contrast $q < 10^3$ for typical molecular vibrations\cite{Kim2018,Blau2018}. This implies that the inherent decay of the internal circuits will be slow enough to be able to simulate all physically relevant dynamics before they start contributing to the dynamics. Therefore, the decay rate of $LC$ oscillators is small enough for simulating the bath spectral densities of realistic physical/chemical systems.

\begin{figure}[htbp]
\includegraphics[width=\columnwidth]{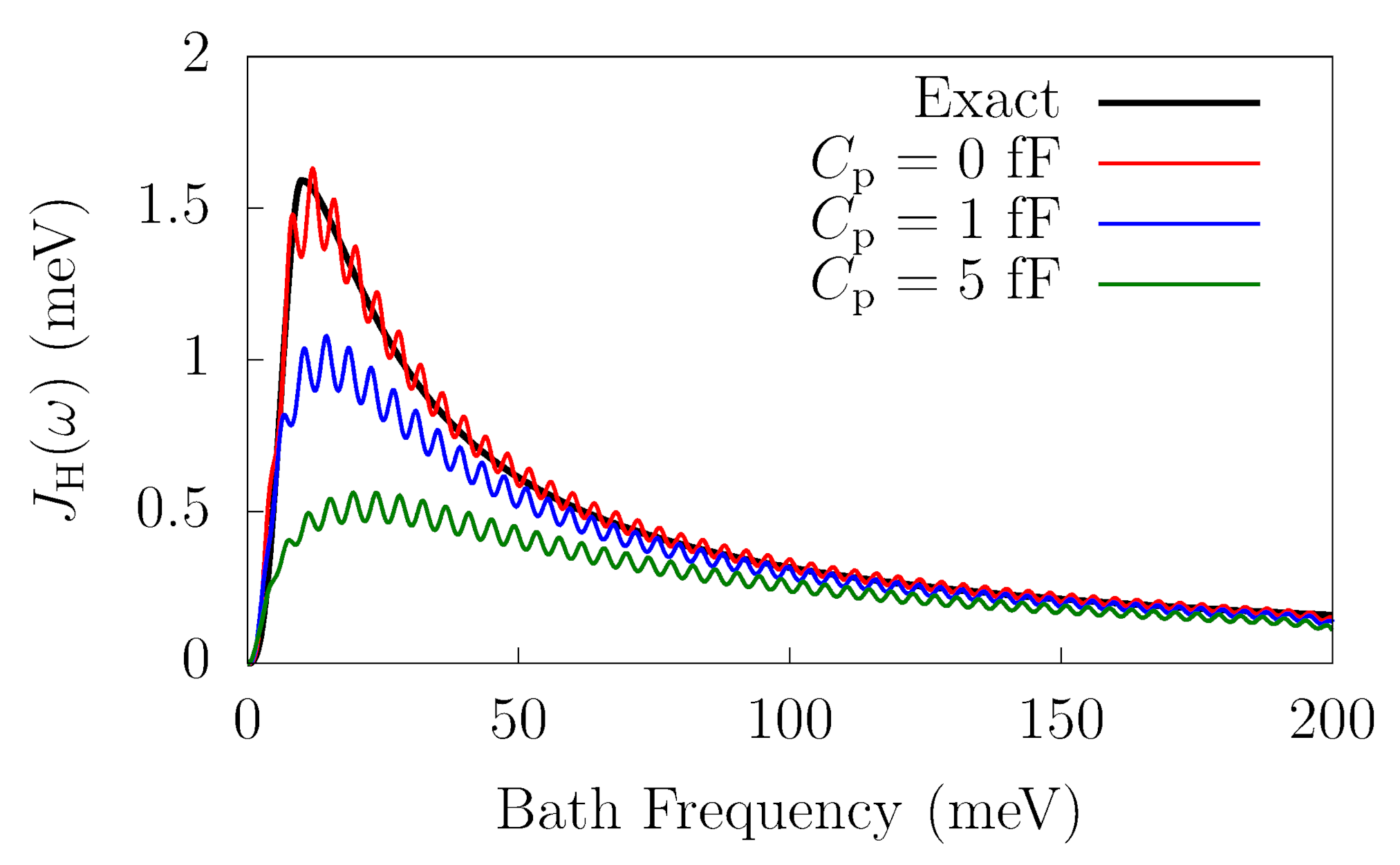}
\caption{\label{fig:para_c} Effect of parasitic capacitance on the bath spectral density realized by the QBS. The target spectral density $J_\ti{H}(\omega)$ (black) was discretized into 50 equally spaced $RLC$ oscillators with a fixed damping parameter of 2 meV (red), and parasitic capacitances of 1 fF (blue) and 5 fF (green) were added to each oscillator.}
\end{figure}

\paragraph{Parasitic capacitance} due to the wire that serially connects individual $LC$ or $RLC$ oscillators can affect the overall impedance of the QBS. This effect can be mimicked by adding an additional parallel capacitance $C_\ti{p}$ to each oscillator, which increases the overall capacitance of the $j$-th oscillator from $C_j$ to $C_j + C_\ti{p}$. The size of $C_\ti{p}$ in microelectronic circuits is often assumed to be a few femtofarads.\cite{Leppaekangas2018} To examine how detrimental this is to our simulator, we discretized $J_\ti{H}(\omega)$ (\stn{subsection:QBS_req}) into 50 $RLC$ oscillators with $\Omega_{j0} = $ (4 meV)$\times j$ and $\Gamma_j = 2$ meV. The effective $J_\ti{H}(\omega)$ of the QBS was then simulated for different values of $C_\ti{p}$ by using \eqs{eq:RLC_SPD}{eq:spd_vs_RLC}. For simulation parameters we have chosen $k_\ti{s} = 0.6$ and $\omega^* = \omega_\ti{c}$, where $\omega^*$ is the cutoff frequency for the splitting function $S(\omega, \omega^*)$ introduced in \stn{subsection:QBS_req}. We have set $\lambda$ and $\omega_\ti{c}$ as the same as in the emulations (\stn{subsection:emulation}). The results are presented in \fig{fig:para_c}, which shows that the parasitic capacitance reduces the intensity of the spectral density by suppressing zero-point voltage fluctuation of the circuits. Such a reduction can be compensated by increasing the impedance at the affected frequencies, which can be achieved by serially connecting multiple identical circuit units even in the presence of the parasitic capacitance.


\section{Possible Extensions}\label{stn:extensions}

We have demonstrated that the dynamics of a two-level open quantum system can be simulated by combining a DQD and a QBS. We now discuss how to extend the present proposal to systems of larger dimensionality and point out some novel extensions that are unique to our setup.

\subsection{Multidimensional Systems}
The open two-level quantum system can be used as a building block for constructing multi-level systems. The most straightforward way to achieve this in our setup is by placing multiple DQD units in head-to-tail configurations, which creates exchange coupling between the electrons in adjacent DQDs. This enables the excitation to migrate across the DQDs, and the rate of migration can be controlled by changing the distance between the DQDs.\cite{Petta2005,Dial2013} We note that an experimental demonstration of such a phenomenon has been already reported for a four-QD device\cite{Kandel2019} and efforts toward the precise manipulation of electrons in increasingly larger arrays of QDs are ongoing.\cite{Kandel2019,Philips2022}

Constructing such a linear chain of DQDs and coupling each unit to a separate QBS will lead to a one-dimensional Holstein model\cite{Holstein1959}, which can simulate the transfer of excitations or charges among molecular aggregates (\fig{fig:scheme}a). Other geometries can be constructed by establishing long-range connections between the DQD units\cite{Harvey2018} or by using QD structures involving three or more QDs\cite{Medford2013,Russ2018}. In addition, simultaneously connecting a QBS to multiple DQD units will enable isolating emerging correlations between subsystems that are mediated by an environment.


\subsection{Non-standard Initial Conditions}
Simulation methods for open quantum system dynamics often rely on factorizable initial conditions $\hat{\rho}(0) = \hat{\rho}_\ti{s}(0) \otimes \hat{\rho}_\ti{b}(0)$ where the bath is at thermal equilibrium [\eq{eq:Bath_DM}]. However, realistic dynamics is often launched from non-thermal bath conditions or correlated system-bath states. Our setup can be used to treat such exotic initial conditions by generalizing the initialization scheme in \stn{stn:mapping2}. Instead of placing the QBS under a constant bias voltage, we can apply time-dependent driving or use nonlinear circuit elements\cite{Hofheinz2009} to engineer non-thermal bath densities. Correlated system-bath states can be also generated by connecting the QBS to QDs prior to the simulation and applying the similar preparation schemes.

Implementing such initialization schemes requires the ability to immediately switch between different Hamiltonians at the start of the simulation. This can be achieved in our setup by harnessing the fast response time ($\sim 10$ ps) of the electrical QD-QBS connection.

\subsection{Quantum Transport}
In addition to systems with a fixed number of electrons, in the QD setup it is possible to also open the boundaries and make the QDs host electron transport by properly tuning the chemical potentials of the source, drain, and the QDs.\cite{Kouwenhoven1997} Connecting the QDs to QBS adds quantum fluctuations to the QD energies, which can be used to simulate the dynamics in Fock space interacting with an external bath. Although the coherent limit of such a situation is relatively well studied,\cite{Datta2005} including the effect of structured quantum noise is beyond the computational reach of state-of-the-art methods such as non-equilibrium Green's function.\cite{Zheng2010,Zhang2013} Analog quantum simulators based on QDs and QBS could naturally handle such challenging cases, which will provide valuable insights on quantum transport phenomena in nanoscale systems such as switching,\cite{Franco2011} rectification,\cite{Han2021}, quantum interference,\cite{Ballmann2012} and controlling the behavior of charge carriers by light.\cite{Boolakee2022}

\section{Conclusions}\label{stn:conclusions}

In conclusion, we have developed a useful strategy for analog quantum simulation of open quantum systems by combining semiconductor QDs and arrays of serially connected $RLC$ circuits. The proposed approach exactly captures the decoherence and spontaneous emission of fully quantum harmonic environments of arbitrary complexity. As such, it goes beyond classical noise models, Markovian or perturbative treatments of system-bath interactions. Further, it has a quantum advantage as the computational time of the simulation does not increase with the complexity of the environment. 

As a specific example, we have established a map between a two-level quantum system embedded in a thermal environment and a DQD connected to a QBS. As a simulation subspace we choose the $S-T_-$ qubit as it simultaneously offers favorable sensitivity to the QBS environment and sufficient long coherence times to enable quantum simulation. The QBS interacts with the QDs electrically through quantum fluctuations of the $RLC$ circuit voltages. The diagonal matrix elements of the system Hamiltonian are experimentally controlled through the detuning and an homogeneous magnetic field; the off-diagonal couplings via magnetic field gradients. By applying voltages to the QBS the spin-boson and displaced oscillator models can be exactly mapped onto the simulator. 

The utility of the map was confirmed by numerically emulating the simulator through HEOM computations of the dynamics that include both simulator and spectator states of QDs. These emulations demonstrate that the quantum dynamics of the simulator can be confined to the desired simulation subspace and that they offer a faithful description of the target dynamics in an experimentally accessible regime. We find that the overall sensitivity of the QDs to the QBS is a useful target simulation parameter that determines the quality of simulation and the experimental requirements in building the platform.

Based on the map and the emulations, we identify the requirements needed to physically build the simulator, including the temperature scaling and sensitivity in the map, tunnel coupling values, magnetic fields, and QBS size, materials and range of frequencies. To reduce the size of the QBS we propose to divide the spectral density into a low frequency and a high frequency component, and capture the effects of the low frequency through classical noise without a significant loss of accuracy. The high-frequency component of the spectral density can be constructed by $1-100$ individual $RLC$ circuits per color.


An advantage of the proposed strategy is that the electrical connection of the QDs with the QBS offers the possibility of conveniently manipulating the importance of individual components in the bath spectral density through voltage dividers containing variable resistors. 
Further, the QBS and QDs can be spatially separated and experience different temperatures, offering additional flexibility in simulator design.

Overall, the proposed setup provides a general strategy for the simulation of open quantum systems, even those of chemical complexity. The strategy can be physically realized with existing technology,\cite{Sundaresan2015,Kollar2019} and is therefore near term. This contrasts with other strategies that are being advanced that require a universal quantum computer. While the quantum hardware operates at cryogenic temperatures, by controlling energy level spacing it can be made to simulate dynamics at arbitrary target temperatures. We have also outlined how to extend the applicability of our setup to multidimensional systems, non-standard initial conditions and quantum transport through open boundary systems. Indeed, recent advances in constructing fully controllable quantum dots with 6-10 individual quantum dot units\cite{Mills2019,Philips2022} open a clear path toward the simulation of many-level systems.

Quantum simulators~\cite{Daley2022} have the power to shed new light on some of the most challenging problems in modern science. The simulation strategy that we have advanced is expected to be of general utility to understand the dynamics of open quantum systems in chemistry, physics and quantum information science.  This proposal represents an exciting emerging direction in quantum simulation and a new promising strategy in the ecosystem of molecular simulation methods.



\begin{acknowledgments}
This work was supported by a PumpPrimer II award of the University of Rochester. IF acknowledges support from the National Science Foundation under grant number CHE-1553939 and CHE-2102386, and the Leonard Mandel Faculty Fellowship of the University of Rochester. CWK is grateful to Al{\'a}n Aspuru-Guzik (University of Toronto) and Joonsuk Huh (Sungkyunkwan University) for their advice on the current status of the field.
\end{acknowledgments}

%

\end{document}